\documentclass[11pt]{wlscirep}
\usepackage[utf8]{inputenc}
\usepackage[T1]{fontenc}
\usepackage{bm}
\usepackage{float}
\usepackage{subcaption}
\usepackage{graphicx}
\usepackage{lineno}
\usepackage[font={small}]{caption}
\usepackage{placeins}

\title{In-situ learning harnessing intrinsic resistive memory variability through Markov Chain Monte Carlo Sampling}
\author[1,*]{Thomas Dalgaty}
\author[1]{Niccolo Castellani}
\author[2,*]{Damien Querlioz}
\author[1,*]{Elisa Vianello}
\affil[1]{Universit\'e Grenoble Alpes, CEA, LETI, 38000, Grenoble, France.}
\affil[2]{Universit\'e Paris-Saclay, CNRS, Centre de Nanosciences et de Nanotechnologies, 91120, Palaiseau, France.}

\affil[*]{e-mail: thomas.dalgaty@cea.fr, damien.querlioz@c2n.upsaclay.fr, elisa.vianello@cea.fr}
\begin{abstract}

Resistive memory technologies promise to be a key component in unlocking the next generation of intelligent in-memory computing systems that can act and learn locally at the edge. However, current approaches to in-memory machine learning focus often on the implementation of models and algorithms which cannot be reconciled with the true, physical properties of resistive memory. Consequently, these properties, in particular cycle-to-cycle conductance variability, are considered as non-idealities that require mitigation. Here by contrast, we embrace these properties by selecting a more appropriate machine learning model and algorithm. We implement a Markov Chain Monte Carlo sampling algorithm within a fabricated array of $16,384$ devices, configured as a Bayesian machine learning model. The algorithm is realised in-situ, by exploiting the devices as random variables from the perspective of their cycle-to-cycle conductance variability. We train experimentally the memory array to perform an illustrative supervised learning task as well as a malignant breast tissue recognition task, achieving an accuracy of $96.3\%$. Then, using a behavioural  model of resistive memory calibrated on array level measurements, we apply the same approach to the Cartpole reinforcement learning task. In all cases our proposed approach outperformed software-based neural network models realised using an equivalent number of memory elements. This result lays a foundation for a new path in-memory machine learning, compatible with the true properties of resistive memory technologies, that can bring localised learning capabilities to intelligent edge computing systems.
\end{abstract}
\maketitle
\begin{document}
\newpage
%\linenumbers
%TC:ignore
%TC:endignore

A tantalising prospect for the future of computing is 
the realisation of standalone systems capable of acting, adapting and learning from new experience
locally at the edge \cite{Shi16} - independent of the cloud. %while respecting tight constraints on energy consumption. 
An embedded medical system for example could adapt its operation depending on the evolution of a patient's state.
The models and algorithms %applied %within the domain 
within machine learning offer %themselves as 
%the tools that will enable such systems. 
the enabling tools for such systems.
However, until recently, little attention has been given to the hardware that underpins their inherent computation. Machine learning models are trained using general purpose hardware which 
inherits from the von Neumann organisation \cite{vonNeumann93}. This entails a spatially separate processing and memory and does not owe itself to energy-efficient training. For example, state of the art performance in machine learning is currently being obtained with neural network models \cite{LeCun15}, featuring a very high number of parameters. 
The energy required to train them can be staggering due to the 
transfer of vast quantities of information between 
memory and processing centres on the hardware \cite{Strubell19,Li16}. 
These demands are not consistent with the energy requirements of edge computing\cite{Shi16}. It is therefore required to abandon the von Neumann paradigm in favour of another, where memory and processing can co-exist at the same location.

\begin{figure}[p]
  \begin{subfigure}[h]{1\linewidth}
  \centering
  \includegraphics[width=0.75\linewidth]{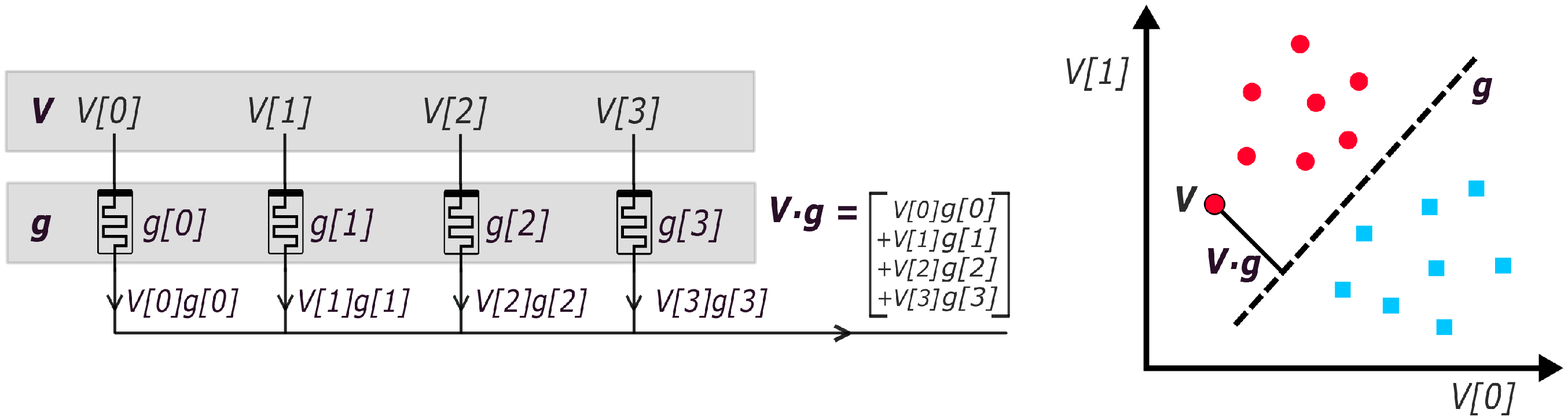}
  \caption{}\label{fig5:dotprod}
  \hfill
  \end{subfigure} 
  \begin{subfigure}[h]{1\linewidth}
    \centering    \includegraphics[width=0.85\linewidth]{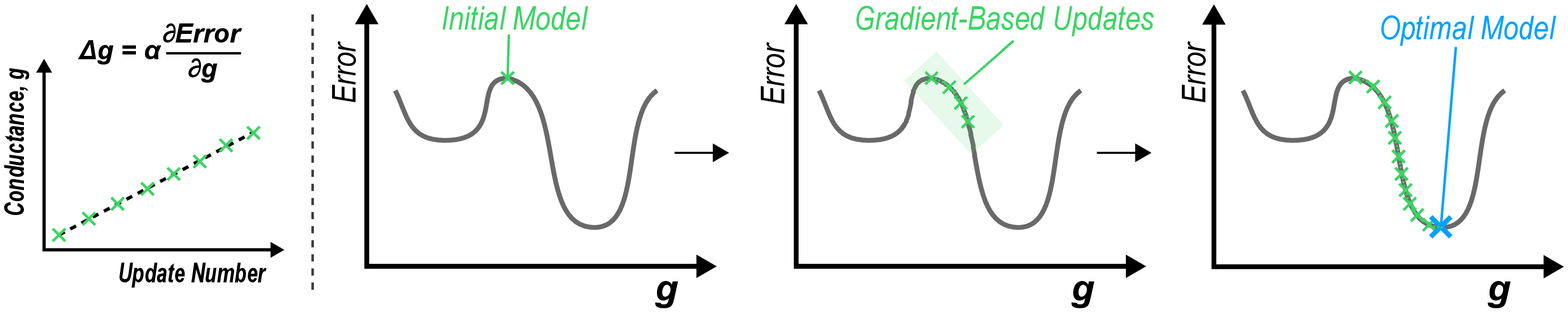}
    \caption{}\label{fig4:gradient}
    \hfill
  \end{subfigure} 
    \begin{subfigure}[h]{1\linewidth}
    \centering    \includegraphics[width=0.85\linewidth]{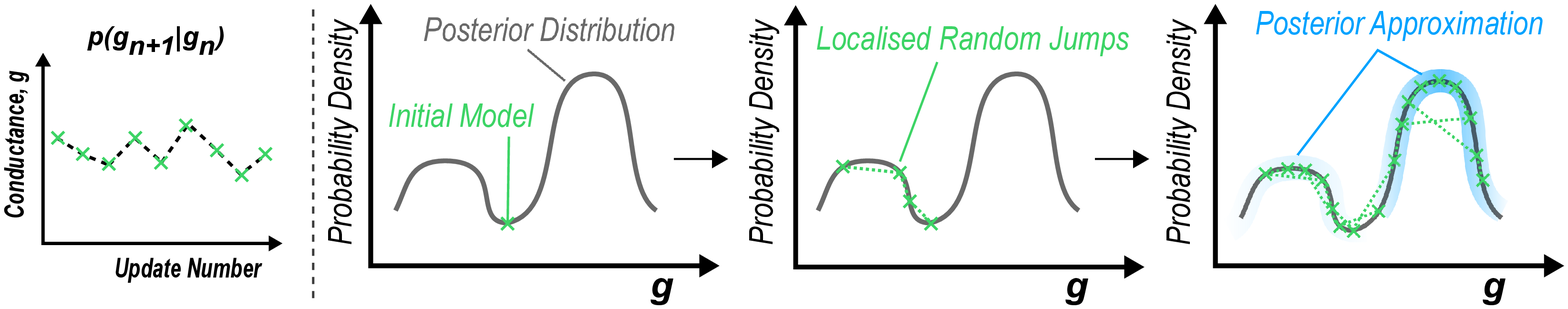}
    \caption{}\label{fig4:sampling}
    \hfill
  \end{subfigure} 
  \caption{
  \textbf{Strategies for training RRAM crossbars.} 
  (\subref{fig5:dotprod}) (left) A conductance model \textbf{g}, composed of four resistive memory elements, defines a linear boundary which 
  (right) separates two classes of data (red circles from blue squares). Through application of a voltage vector \textbf{V} to the top electrode of the parallel resistive memories, the summed current flowing out of the common node at the the bottom electrode is equivalent to the dot-product $\textbf{V}\cdot~\textbf{g}$,
  which can then be used to determine to what class the data point \textbf{V} belongs.
  (\subref{fig4:gradient}) (left) Gradient-based learning algorithms iteratively compute the derivative of an error metric with respect to a conductance model $\textbf{g}$, multiplied by a learning rate $\alpha$, to determine updates to be applied to the $g$ parameters. The ideal RRAM device should be capable of high precision and linear conductance updates. (right) The three panels show the gradient-descent algorithm for an increasing number of model updates (green crosses). From an initial model the algorithm performs gradient-based updates until it converges to a local minimum in error.
  (\subref{fig4:sampling}) (left) Sampling algorithms use a proposal distribution ($p(g\textsubscript{n+1}|g\textsubscript{n})$) to propose random updates to model conductance parameters which are then either accepted or rejected. The ideal RRAM device for sampling algorithms should offer random conductance updates deriving from a known probability distribution. (right) The three panels illustrate how a sampling algorithm performs local random jumps on the posterior distribution for an increasing number of sampling operations. From an initial model, a proposal distribution generates a series of localised random jumps (dashed green lines) which are then either accepted (green cross) or rejected. The algorithm tends to accept models of a higher probability density on the posterior distribution. After a sufficient number of iterations the accepted models can be used together as an approximation of the posterior distribution (blue haze).}
  \label{fig4:gradvssamp}
\end{figure}

Resistive random access memory (RRAM) technologies, often referred to as memristors \cite{Chua71,Strukov08}, hold fantastic promise for realising such in-memory computing systems, owing to their extremely efficient implementation of the dot-product (or
multiply-and-accumulate)
operation that pervades machine learning - relying simply on 
Ohm's law \cite{Xia19,Ambrogio18,Prezioso15,Wang19,Miao18,Li18} (Fig.~\ref{fig5:dotprod}).
These technologies come in many flavours \cite{Wong10,Chappert07,Liu12,Beck00}, and intense effort is currently directed towards their use as synaptic elements in hardware neural networks for edge computing systems \cite{Xia19,Ambrogio18,Prezioso15,Wang19,Miao18,Li18}.
Currently, approaches for training such systems in-situ
revolve around in-memory implementations of gradient-descent with the back-propagation algorithm \cite{Ambrogio18,Gokmen17,Wang19,Miao18,Li18}.
However, implementing back-propagation in such hardware remains a formidable challenge due to multi-fold non-ideal device properties: 
non-linear conductance modulation \cite{Burr15}, lack of stable multi-level conductance states \cite{Garbin15,Sebastian15}, 
as well as 
device
variability \cite{Guan12,Yu12pt2}. 
Considering these real device properties, the performance of systems 
can be lower than that obtained on conventional %von Neumann graphical processing units (GPUs)
computing systems
\cite{Sidler16,Bennett19,Agarwal16}. Several non-ideality mitigation techniques \cite{Ambrogio18,Gokmen17,Nandakumar18,Miao18,Li18,Boybat17} 
enhance system accuracy but ultimately curtail the efficiency of the in-memory computing approach. 
On the contrary, approaches based on neuroscience-inspired learning algorithms, such as spike-timing dependent 
plasticity feature resilience, and in fact sometimes benefit from device non-idealities \cite{Serb16,Querlioz13,Dalgaty19}. 
However, these brain-inspired models cannot yet match state-of-the-art machine learning models 
when applied to practical tasks. 
Further research has taken a bolder stance on the issue of resistive memory non-idealities and instead propose that they should be actively embraced.
For example, the cycle-to-cycle conductance state variability, which has been leveraged as a source of entropy in random number generation \cite{Chen15,Balatti15}, has also been exploited in stochastic artificial intelligence algorithms such as Bayesian reasoning \cite{vodenicarevic2017low,shim2017stochastic,faria2018implementing}, population coding neural networks \cite{mizrahi2018neural} and in-memory optimisation \cite{camsari2017stochastic,borders2019integer}.
Unfortunately,
these approaches sacrifice the key property of conductance non-volatility - the basis of resistive memory's potential for efficient in-memory computing. 
%Further, the performance of these approaches suffer from device-to-device variability that in turn also require mitigation through calibration \cite{borders2019integer} or whitening \cite{vodenicarevic2017low}.

In this work, we present an alternative approach which simultaneously exploits conductance variability and conductance non-volatility without requiring mitigation of other device non-idealities.
From the perspective of cycle-to-cycle conductance variability, we propose that resistive memories can be viewed as physical random variables which can be exploited to implement in-memory 
Markov Chain Monte Carlo (MCMC) sampling algorithms\cite{Hastings70}. 
We show how a resistive memory based Metropolis-Hastings MCMC sampling approach can be used to train, in-situ, a Bayesian machine learning model realised in an array of resistive memories. Crucially, the devices that perform the critical sampling operations are also those which store the parameters of the Bayesian model in their non-volatile conductance states. This eliminates the need to transport information between processing and memory and instead relies on the physical responses of nanoscale devices under application of voltage pulses inside of the memory structure itself. 

In order to demonstrate the practicality of RRAM-based MCMC sampling, we implemented an experimental system consisting of a computer-in-the-loop with a fabricated array of $16,384$ resistive memory devices which are configured as a Bayesian machine learning model. The computer is responsible for configuring voltage waveforms which iteratively read and reprogram the conductance states of devices in order to train the RRAM-based model in-situ. In a first experimental  realisation, we train the system to solve an illustrative classification task and in a second we apply it to the detection of malignant breast tissue samples. Finally, we extend the approach, through a behavioural simulation calibrated on array level measurements, to the Cartpole reinforcement learning task. We benchmark the resulting performance against software-based neural network models, realised with an equivalent number of memory elements, and find that our proposed approach, even in the presence of unmitigated device-to-device variability, is able to outperform both of these benchmarks.

%TC:ignore
%TC:endignore
%\clearpage

\section*{Resistive Memory Based MCMC Sampling}

Arrays of resistive memory devices 
%(RRAM)
are capable of implementing extremely efficient in-memory machine learning models \cite{Xia19,Ambrogio18,Prezioso15,Wang19,Miao18,Li18}.
Considering an RRAM-based logistic regression classifier, one of the most canonical models in machine learning, the circuit of $M$ parallel devices shown in Fig.~\ref{fig5:dotprod} defines a hyper-plane that separates two classes of data. Each parameter of the logistic regression model is defined by the conductance of one of the $M$ devices. The response of this conductance model, $\textbf{g}$, can be inferred by presenting a voltage vector $\textbf{V}$, encoding a data point, to the top terminals of the devices and sensing the current that flows out of the common, bottom node. This current evaluates the dot-product between the two vectors ($\textbf{V}\cdot~\textbf{g}$) in the output current.

The dominant approach in RRAM-based machine learning for training the conductance parameters of a model is gradient-based optimisation whereby an error, or cost function, is iteratively differentiated with respect to the current parameters of the model. The resulting derivative provides precise conductance updates which, after being applied over a sufficient number of iterations, guide the model down the slope of an error gradient such that it settles into a minimum (Fig.~\ref{fig4:gradient}). 
This results in a locally-optimal model that can then be applied to tasks through inference. However, performing this type of training in-memory is extremely challenging as resistive memory technologies feature highly non-linear and variable conductance updates that do not naturally offer the high precision required by this class of algorithm \cite{Burr15,Garbin15,Sebastian15,Guan12,Yu12pt2}. 
Furthermore, in a deterministic model (Fig.~\ref{fig5:dotprod}), each parameter is described by a single value, and it is not possible to account for parameter uncertainty.
Describing uncertainty in parameters is important when dealing with small datasets, noisy sensors or representing confidence in a prediction for a safety-critical edge application \cite{Ghahramani15} - an embedded medical system for example.
To account for uncertainty, it is preferable to construct a Bayesian model. In this case, parameters are represented, not by single values, but by probability distributions. 
The distribution of all of the parameters in a Bayesian model, given observed data, is called the posterior distribution, or simply the posterior. 
As an analogue to deriving an optimal deterministic model through gradient-based updates, the objective in Bayesian machine learning is to learn an approximation of the posterior distribution. When the posterior has been approximated it can be applied to tasks through model inference. To approximate the posterior, sampling algorithms, most commonly Markov Chain Monte Carlo (MCMC) sampling \cite{Hastings70}, are employed. 
Instead of descending an error gradient, MCMC sampling algorithms make localised random jumps on the posterior distribution and continuously store models that lie on it (Fig.~\ref{fig4:sampling}).
The algorithm jumps from a current location in the model space $\textbf{g}$ to a proposed location $\textbf{g}\textsubscript{p}$, according to a proposal distribution $p(\textbf{g}\textsubscript{p}|\textbf{g})$ which is usually a normal random variable. % $\mathcal{N}(\textbf{g},\sigma)$. 
By comparing the proposed and current models, a decision is made on whether to accept or reject the proposed model. If accepted, the next random jump is made from the recently accepted model.
MCMC sampling algorithms are configured to accept more samples from regions of high probability density on the posterior that would correspond to a lower error in the gradient-based case of Fig.~\ref{fig4:gradient}. After a sufficiently large number of such iterations, the accepted samples can be used together as an approximation of the posterior distribution.

%TC:ignore
\begin{figure}[p]
  \centering
  \begin{subfigure}[h]{0.25\linewidth}
    \centering
    \includegraphics[width=1\linewidth]{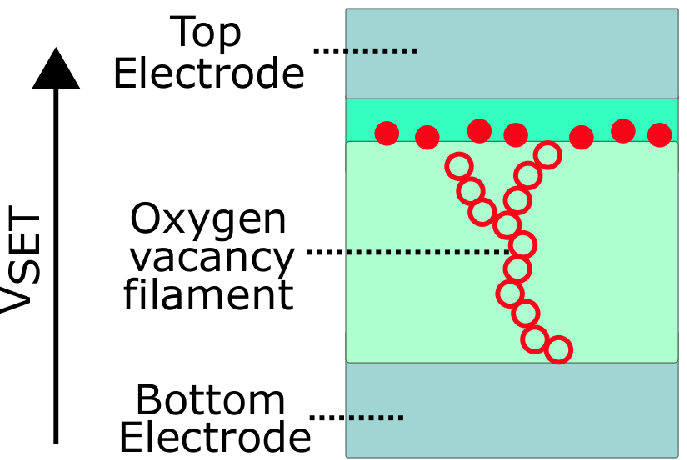}%
    \caption{}\label{fig1:oxramset}
    \hfill
  \end{subfigure}  
  ~
  \begin{subfigure}[h]{0.25\linewidth}
    \centering
    \includegraphics[width=1\linewidth]{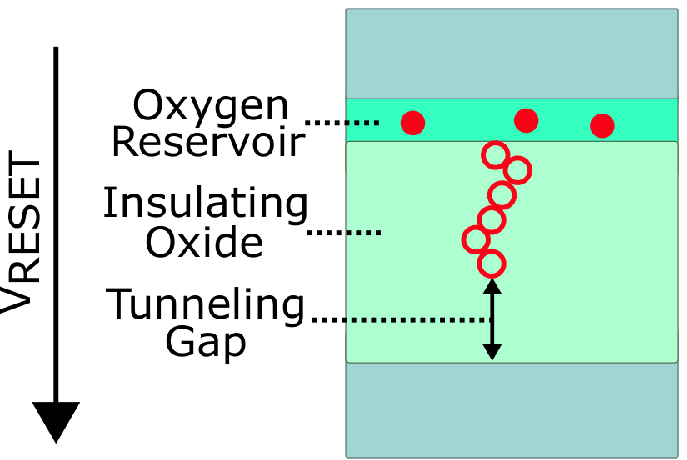}%
    \caption{}\label{fig1:oxramreset}
    \hfill
  \end{subfigure} 
  ~
  
  \begin{subfigure}[h]{0.31\linewidth}
    \centering
    \includegraphics[width=1\linewidth]{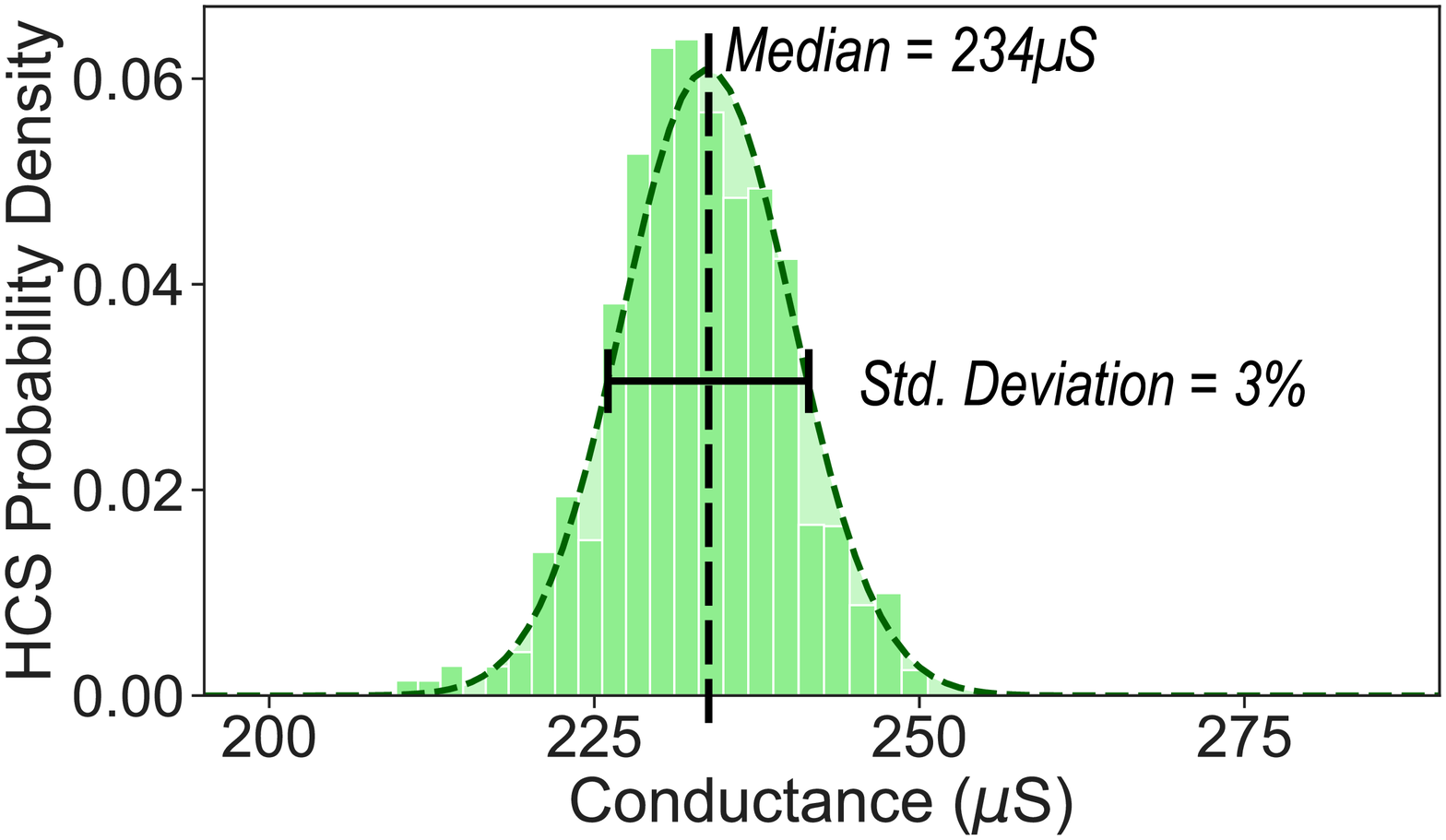}%
    \caption{}\label{fig1:hcscycletocycle}
    \hfill
  \end{subfigure}
  ~
  \begin{subfigure}[h]{0.33\linewidth}
    \centering
    \includegraphics[width=1\linewidth]{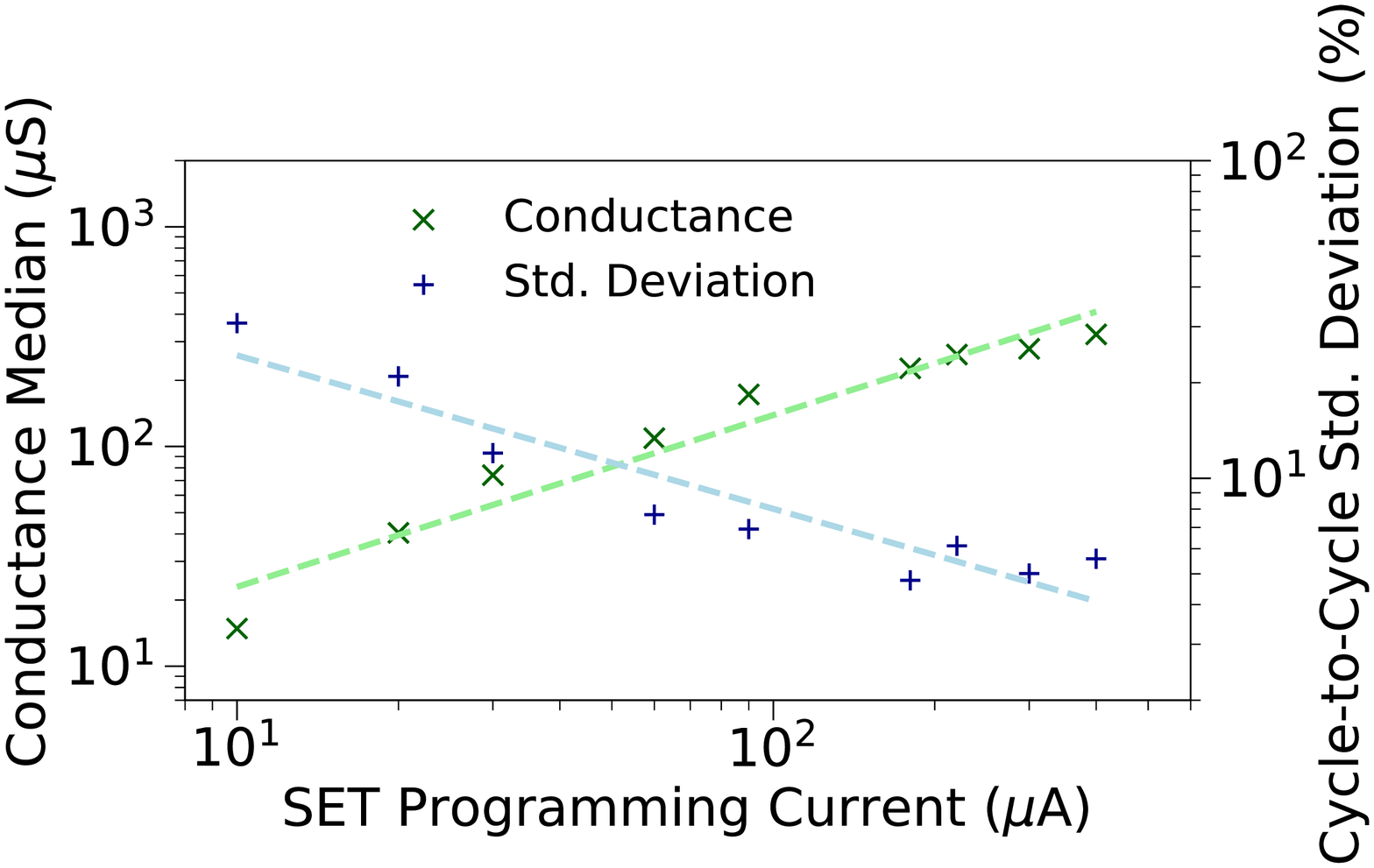}%
    \caption{}\label{fig1:hcspowerlaw}
    \hfill
  \end{subfigure}  
  ~
  \begin{subfigure}[h]{0.3\linewidth}
    \centering
    \includegraphics[width=1\linewidth]{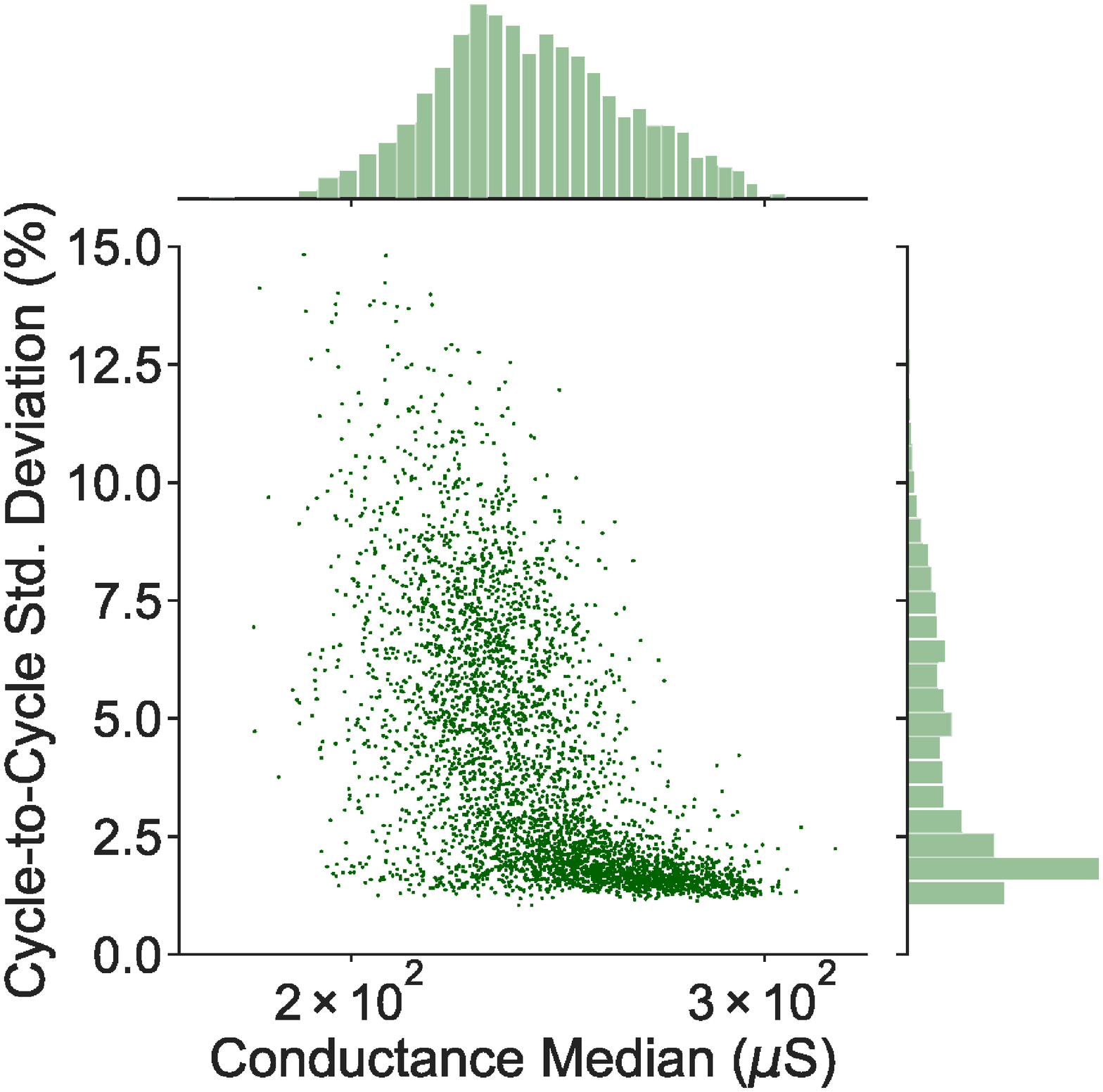}%
    \caption{}\label{fig1:hcsdevicetodevice}
    \hfill
  \end{subfigure} 
  ~
%\end{figure}
%\begin{figure}[p]
%\ContinuedFloat
  \caption{
  \textbf{Electrical characterization of OxRAM cycle-to-cycle and device-to-device variability.} 
  (\subref{fig1:oxramset}) OxRAM device in the high conductance state (HCS). After applying a positive SET voltage waveform %(V\textsubscript{SET}) 
  over the top and bottom electrodes 
  a filament of conductive oxygen vacancies
  %(open red circles) 
  form in the 
  %insulating 
  oxide. %between the two electrodes.
  (\subref{fig1:oxramreset}) OxRAM device in the low conductance state (LCS). After applying a negative RESET voltage waveform, oxygen ions are re-introduced into the oxide from the inter-facial oxygen reservoir between the oxide and top electrode, rupturing the conductive filament. %resulting in a low device conductivity due to the tunneling gap that now exists between the conductive filament and the bottom electrode. %, 
  (\subref{fig1:hcscycletocycle}) Probability density of the HCS cycle-to-cycle 
  %conductance 
  variability for a single OxRAM device, measured over 500 RESET/SET cycles (see Methods), and fitted with a normal distribution (dashed line). 
  %The conductive states, read after each SET operation (see Methods) over 500 are plotted in a histogram and fit using a normal distribution (dashed line). 
  %The normal distribution has a median of 234$\mu$S and a standard deviation of 3\%.
  (\subref{fig1:hcspowerlaw}) 
  %Mean relationship between 
  Cycle-to-cycle conductance median and standard deviation for a population of $4,096$ devices, for a range of SET programming current (see Methods). %Green cross symbols: median HCS conductance as a function of SET programming current, fit using a power law (green dashed line). Blue plus symbols: cycle-to-cycle standard deviation of the HCS over the same range of SET programming currents, also fit with a power law (blue dashed line).
  Both curves are fitted with a power law.
  (\subref{fig1:hcsdevicetodevice}) 
  %Cycle-to-cycle HCS conductance median and standard deviation values for a population of $4,096$ devices. 
  In order to plot this graph,
  $4,096$ devices have been RESET/SET cycled $500$ times under the same programming conditions (see Methods), and the resulting median conductance and standard deviation of each device has been plotted as a single green point - illustrating the device-to-device variability within a population. Two histograms on opposing axes show the probability densities for the conductance median and standard deviation independently. 
  }
\end{figure}
%TC:endignore
%\clearpage

In this paper we realise that, in stark contrast to the case of gradient-based learning algorithms, the properties of resistive memories are incredibly well suited to the requirements of sampling algorithms. This is because the cycle-to-cycle conductance variability, inherent to device programming, is not a nuisance to be mitigated, but instead, a computational resource that can be leveraged by viewing resistive memory devices as physical random variables.
More specifically, we exploit the random variable available in a
hafnium dioxide-based filamentary random access memory \cite{Beck00} (OxRAM), 
co-integrated at array level into a 130~nm complementary metal oxide semiconductor (CMOS) process \cite{Grossi16}. Each memory point in the array is connected in series to an n-type transistor (see Methods). The conductivity of an OxRAM device can be modified by the application of voltage waveforms which, through oxidation-reduction reactions at an interfacial oxygen reservoir between the oxide and top electrode, create or rupture a conductive oxygen-vacancy filament between the electrodes. The device can be SET into a high conductive state (HCS), by applying a positive voltage across the device (Fig~\ref{fig1:oxramset}) and thereafter RESET into a low conductive state (LCS), by applying a negative voltage across the device (Fig~\ref{fig1:oxramreset}). 
Each time the device is programmed a unique HCS conductance is achieved - resulting from the random distribution of oxygen-vacancies within the oxide\cite{Guan12,Yu12pt2}. If this conductance is measured over successive cycles, a normally distributed cycle-to-cycle conductance probability density emerges (Fig.~\ref{fig1:hcscycletocycle}). The SET operation is therefore analogous to drawing a random sample from a normal distribution. In addition, the median conductance of this probability distribution can be controlled by limiting the SET programming current ($I\textsubscript{SET}$) via the gate-source voltage of the series transistor. The relationship between the conductance median and SET programming current follows a power law\cite{Ielmini11} and the standard deviation of the distribution also depends on the SET programming current (Fig.~\ref{fig1:hcspowerlaw}). Therefore, manifested in the physical response of these nanoscale devices, we find the essential computational ingredient required to implement in-memory MCMC sampling algorithms - a physical normal random variable - 
which can be exploited to propose new models based on a previous one. 

%TC:ignore
\begin{figure}[p]
  \centering
  \begin{subfigure}[h]{0.45\linewidth}
    \centering
    \includegraphics[width=1\linewidth]{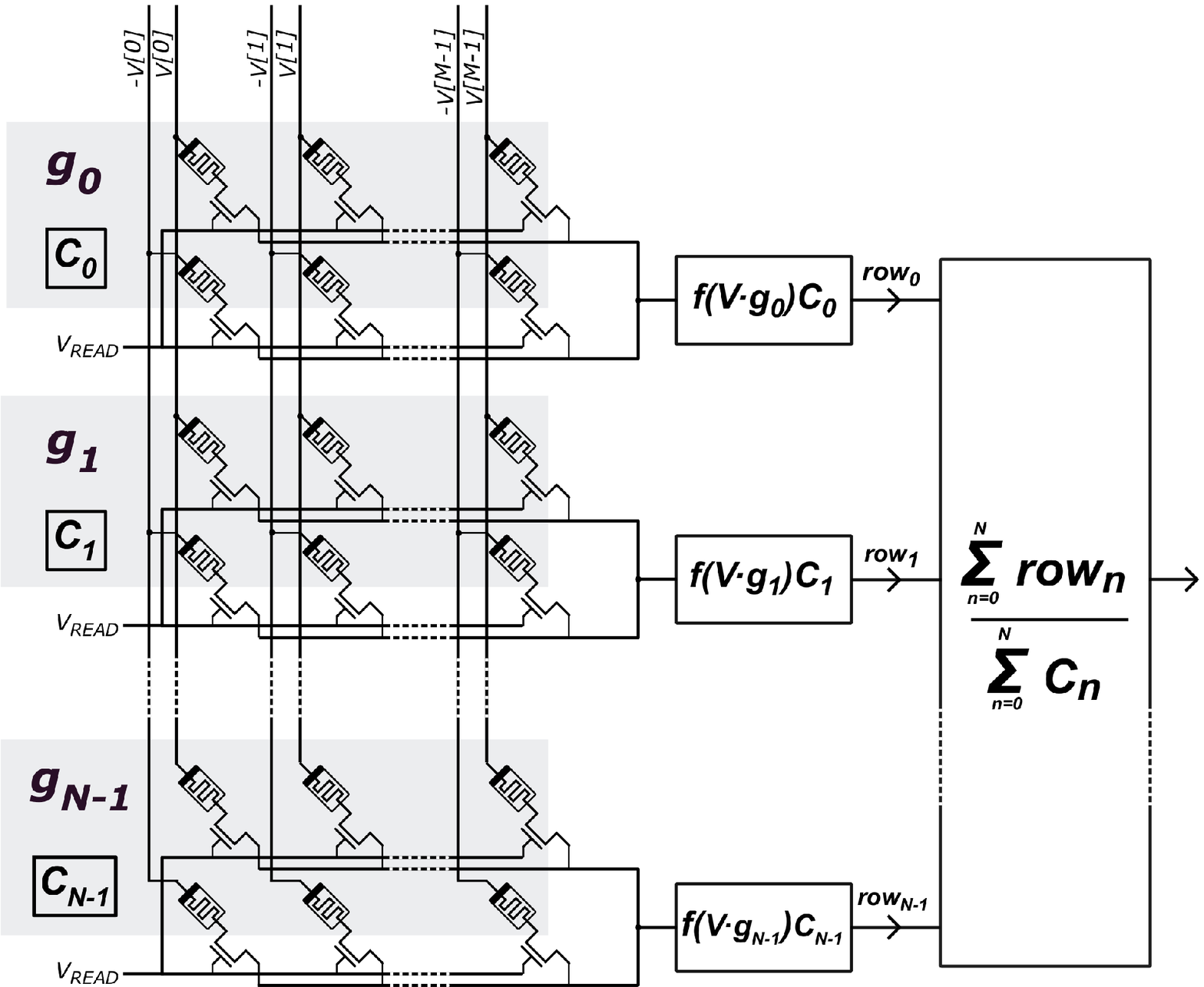}%
    \caption{}\label{fig2:bayesianmodel}
    \hfill
  \end{subfigure}
  ~
  \begin{subfigure}[h]{0.4\linewidth}
    \centering
    \includegraphics[width=1\linewidth]{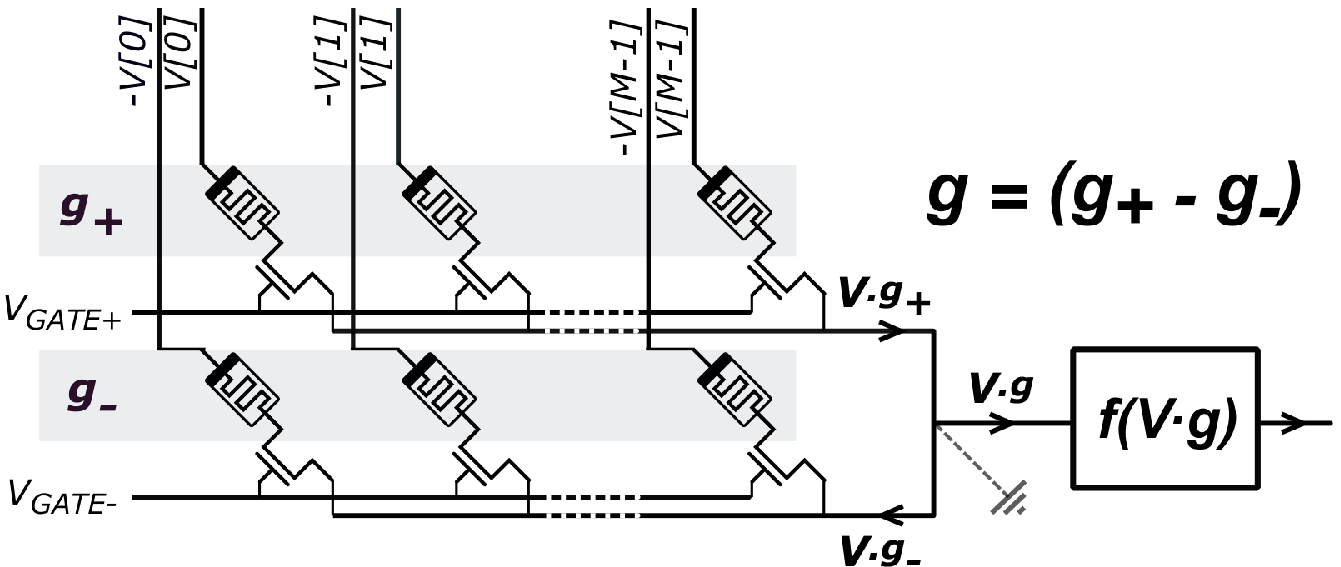}%
    \caption{}\label{fig2:diffrow}
    \hfill
  \end{subfigure}
  ~
  \begin{subfigure}[h]{0.4\linewidth}
    \centering
    \includegraphics[width=1\linewidth]{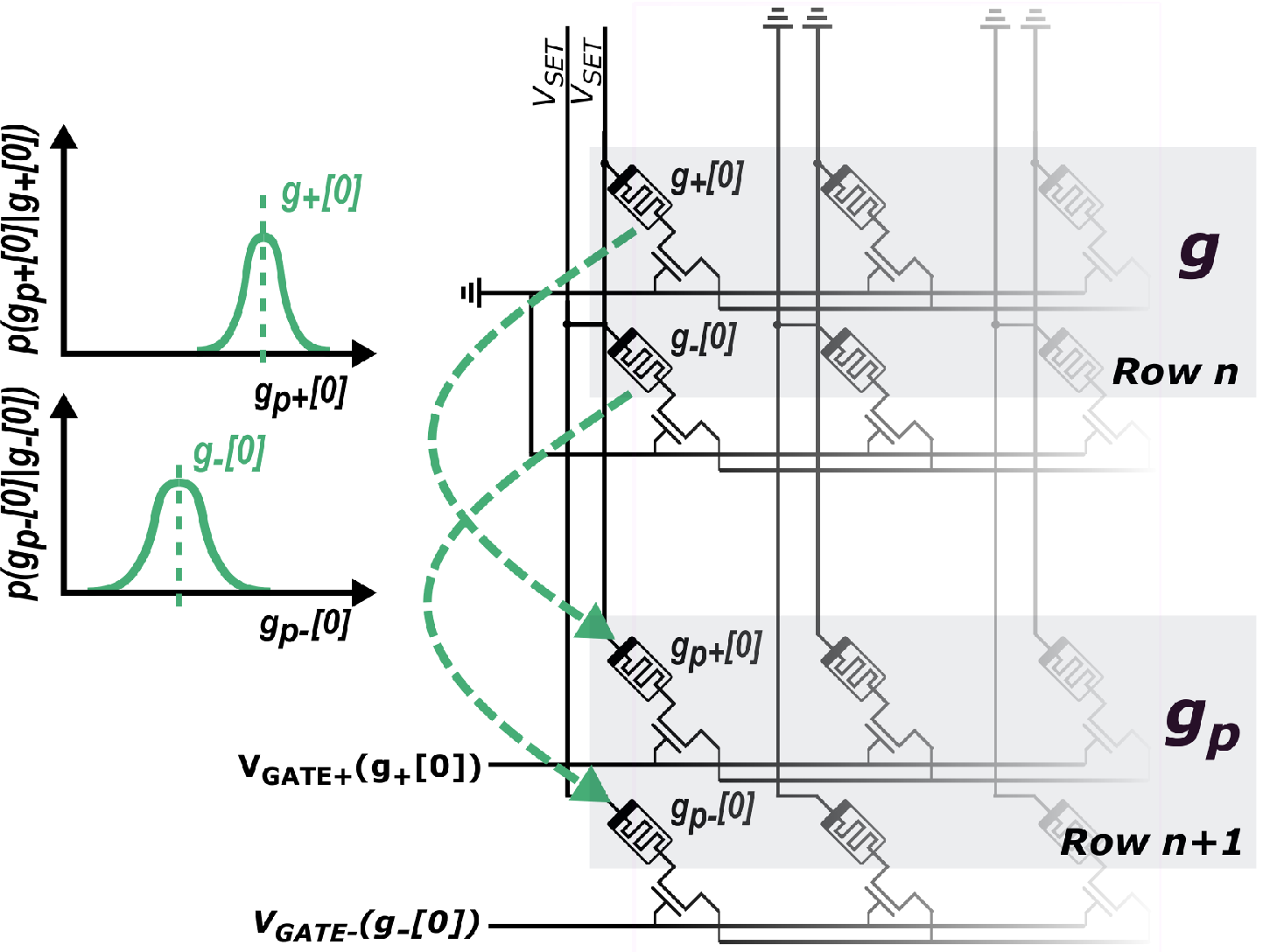}%
    \caption{}\label{fig2:proposal}
    \hfill
  \end{subfigure}
  ~
  \begin{subfigure}[h]{0.5\linewidth}
    \centering
    \includegraphics[width=1\linewidth]{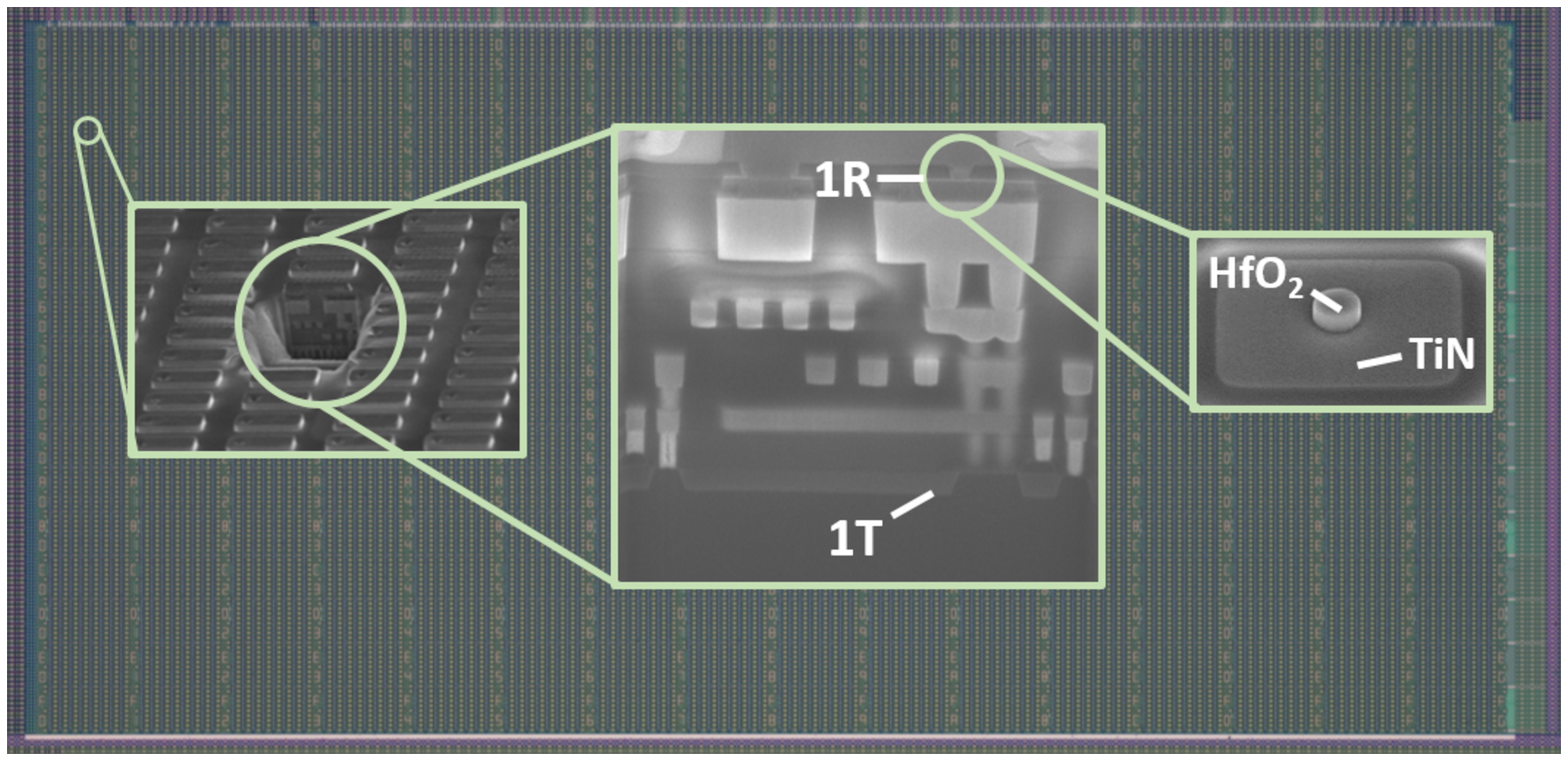}%
    \caption{}\label{fig2:arrayimage}
    \hfill
  \end{subfigure}
  \caption{
  \textbf{Implementation of Metropolis-Hastings Markov Chain Monte Carlo (MCMC) on a fabricated RRAM array.} 
  (\subref{fig2:bayesianmodel}) Memory array architecture where RRAM conductances store the posterior approximation and calculation used to perform inference with a learned posterior. Each of the $N$ rows store a single conductance model $\textbf{g}\textsubscript{n}$ and features a digital counter element, $C\textsubscript{n}$. 
  %During inference, a data point $\textbf{V}$ is applied across the array columns and the dot-product between it and each model is inserted into a function - evaluating $f(\textbf{V}\cdot~\textbf{g}\textsubscript{n})$ at every row. 
  %The output of this function
  %, $row\textsubscript{n}$, 
  %is multiplied by the corresponding row counter value, $C\textsubscript{n}$, and then used in a final calculation where the sum of all row responses is divided by the sum of all counter values in the array.
  (\subref{fig2:diffrow}) A single array row is the differential conductance between conductance vectors $\textbf{g}\textsubscript{+}$ and $\textbf{g}\textsubscript{-}$. A positive voltage vector, $\textbf{V}$, is applied over the top electrodes of $\textbf{g}\textsubscript{+}$ and a negative opposite voltage vector $-\textbf{V}$ is applied over the top electrodes of $\textbf{g}\textsubscript{-}$. If the common, bottom node is pinned at a virtual ground, the current flowing into the function block is equal to the dot-product $\textbf{V}\cdot~\textbf{g}$.
  (\subref{fig2:proposal}) Model proposal step between two array rows. 
  %The conductances $g[0]\textsubscript{+}$ and $g[0]\textsubscript{-}$ in the $0\textsuperscript{th}$ column of the $n\textsuperscript{th}$ row, which contains the current model $\textbf{g}$, are read. 
  Using the known relationship between SET programming current and the read conductances in row $n$ (see Methods), devices (pointed to by the arrows) $g[0]\textsubscript{+}$ and $g[0]\textsubscript{-}$ in the $n+1\textsuperscript{th}$ array row are SET. Their SET programming currents are proportional to the conductances of the corresponding devices in the $n\textsuperscript{th}$ row. 
  The new conductance values of $g[0]\textsubscript{+}$ and $g[0]\textsubscript{-}$ in the $n+1\textsuperscript{th}$ row are thereby sampled from normal random variables with medians equal to the conductances of the devices in the $n\textsuperscript{th}$ array row (the green probability distributions, left). 
  The SET programming currents are fixed by applying the appropriate voltages $V\textsubscript{GATE+}$ and $V\textsubscript{GATE-}$ to the transistor gates of row $n+1$.
  (\subref{fig2:arrayimage}) OxRAM array used in the experiments (see Methods). An optical microscopy image of the fabricated array is shown in the background. Scanning electron microscopy images are superimposed on top. (left) A focused ion beam a etch reveals the cross-section of a 1T1R structure (centre). In the front-end-of-line a transistor (1T) acts as a selector for the OxRAM device (1R) integrated in the back-end-of-line. (right) Imaged before deposition of the top electrode titanium layer, a 10~nm thick, 300~nm wide mesa of HfO\textsubscript{2} rests upon a $TiN$ bottom electrode.
  }
\end{figure}
%TC:endignore
%\clearpage

We propose that the $N \times M$ resistive memory array depicted in Fig.~\ref{fig2:bayesianmodel} can be trained through MCMC sampling and then store, in the distribution of its non-volatile conductance states, the resulting posterior of a Bayesian model.
A single deterministic model, $\textbf{g}\textsubscript{n}$, is stored in each of the rows where its parameters are encoded by the conductance difference between positive $\textbf{g+}\textsubscript{n}$ and negative $\textbf{g-}\textsubscript{n}$ sets of devices - allowing for each parameter to be either positive or negative (Fig.~\ref{fig2:diffrow}). 
The principle of our approach is to generate at each row a proposed model, based on the model in the previous row, inline with the Metropolis-Hastings MCMC sampling algorithm\cite{Hastings70} (see Methods for a detailed description).
% This can be achieved naturally using the gate voltage of the series selector transistor which allows for control of the SET programming current (Fig,~\ref{fig1:hcspowerlaw}) that determines the median of the normal random variable that the new device conductances are sampled from.
Each parameter of the proposed model can be generated naturally, using the OxRAM physical random variable.
% This is achieved through performing a SET operation on each device in the row with a programming current proportional to the conductances of the corresponding devices in the previous row, by applying the appropriate gate voltage to the selector transistor - depicted in Fig.~\ref{fig2:proposal}.
This is achieved through performing a SET operation on each device in the row with a programming current that samples a new conductance value from a normal distribution centred on that of the corresponding device in the previous row
% , by applying the appropriate gate voltage to the selector transistor
(Fig.~\ref{fig2:proposal}).
By computing a quantity called the acceptance ratio (see Methods), a decision is made on whether this proposed model should be accepted or rejected. 
If rejected, the row is reprogrammed under the same conditions - thereby generating a new proposed model.
Additionally, the value of a digital counter, $C_n$, which is associated with the previous row is incremented by one. By tracking the number of rejections in this manner, the contribution of the model in each row to the probability density at each location on the posterior approximation (Fig.~\ref{fig4:sampling}) is accounted for.
If the proposed model is instead accepted, the process is repeated at the next row, and so on until the algorithm arrives at the final row of the array.

At this point the distribution of programmed differential conductances in each of the array columns corresponds to the learned distributions of each of the Bayesian model parameters - the posterior distribution. 
The first rows of the model should however be discarded, as their state is dependent on the initialisation of the devices in zeroth row - an effect referred to as ``burn-in'' in the Metropolis-Hastings algorithm.
After training, the learned posterior distribution in the array can be applied to a task through inference.
%weighted by their corresponding counter values, store an approximation of the posterior distribution. 
%Hereafter, this learned approximation, can be applied to perform tasks through model inference. 
In inference, a new input $\textbf{V}$ is presented to the array whereby the response of all rows, $f(\textbf{V}\cdot~\textbf{g}\textsubscript{n})$, are multiplied by their row counter values and summed. This sum is then divided by the sum of all row counter values as in Fig.~\ref{fig2:bayesianmodel}. The function $f$ depends on the formulation of the machine learning model. 

\begin{figure}[!h]
  \begin{subfigure}{0.32\linewidth}
    \centering
    \includegraphics[width=1\linewidth]{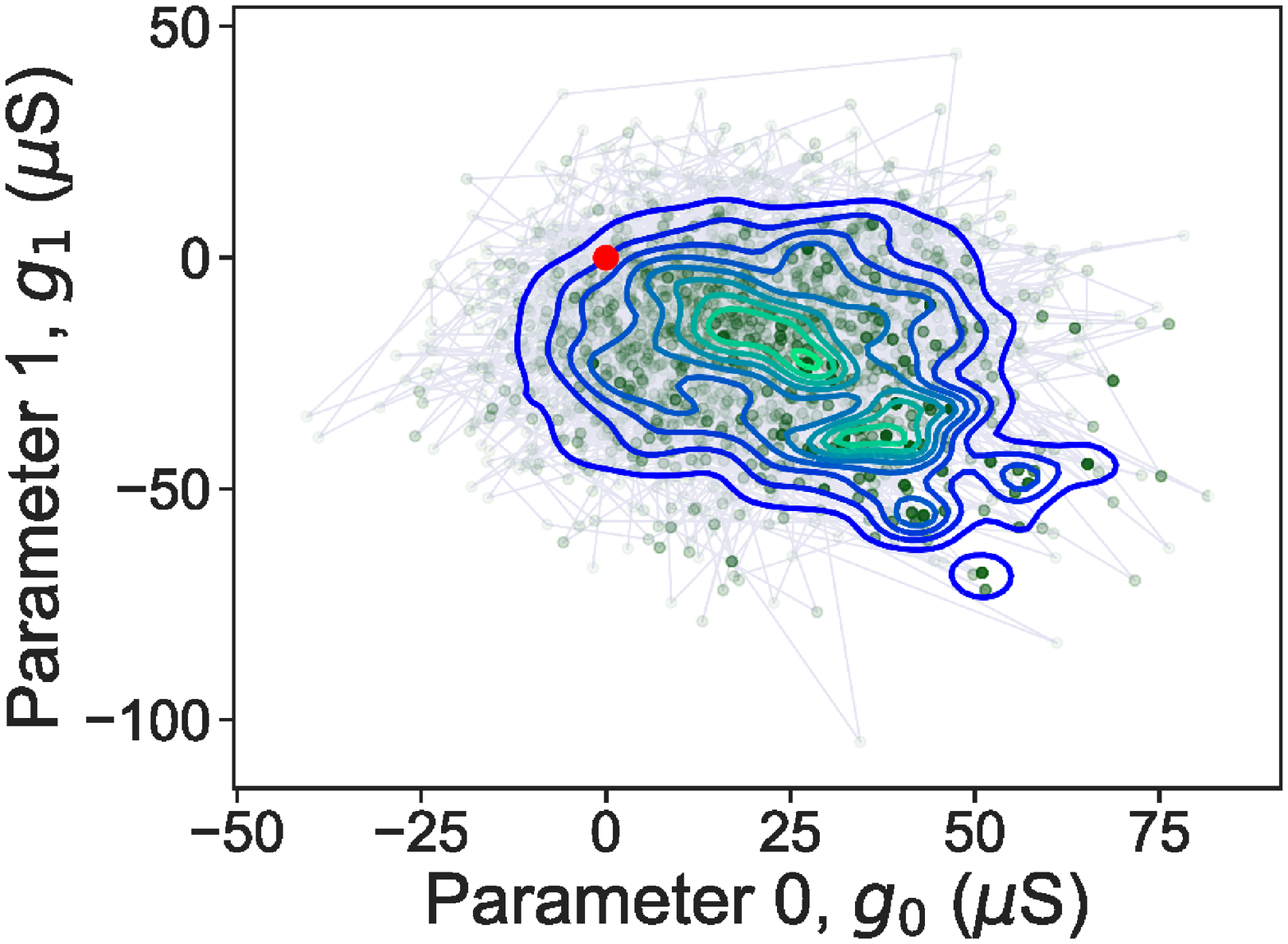}%
    \caption{}\label{fig3:randomwalk}
    \hfill
  \end{subfigure}
  ~
  \begin{subfigure}{0.32\linewidth}
    \centering
    \includegraphics[width=1\linewidth]{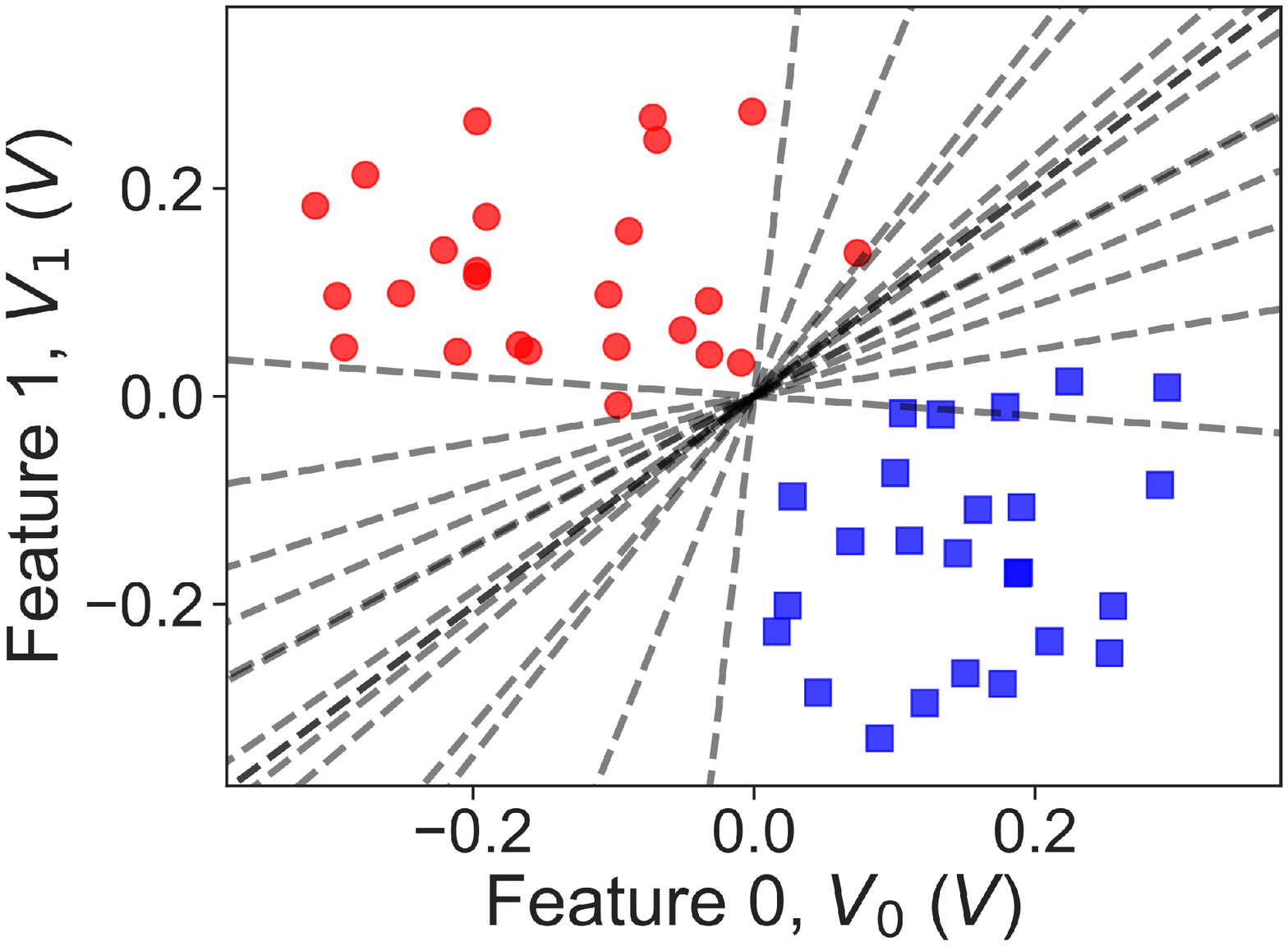}%
    \caption{}\label{fig3:twoclasshyperplanes}
    \hfill
  \end{subfigure}
  ~
  \begin{subfigure}{0.32\linewidth}
    \centering
    \includegraphics[width=1\linewidth]{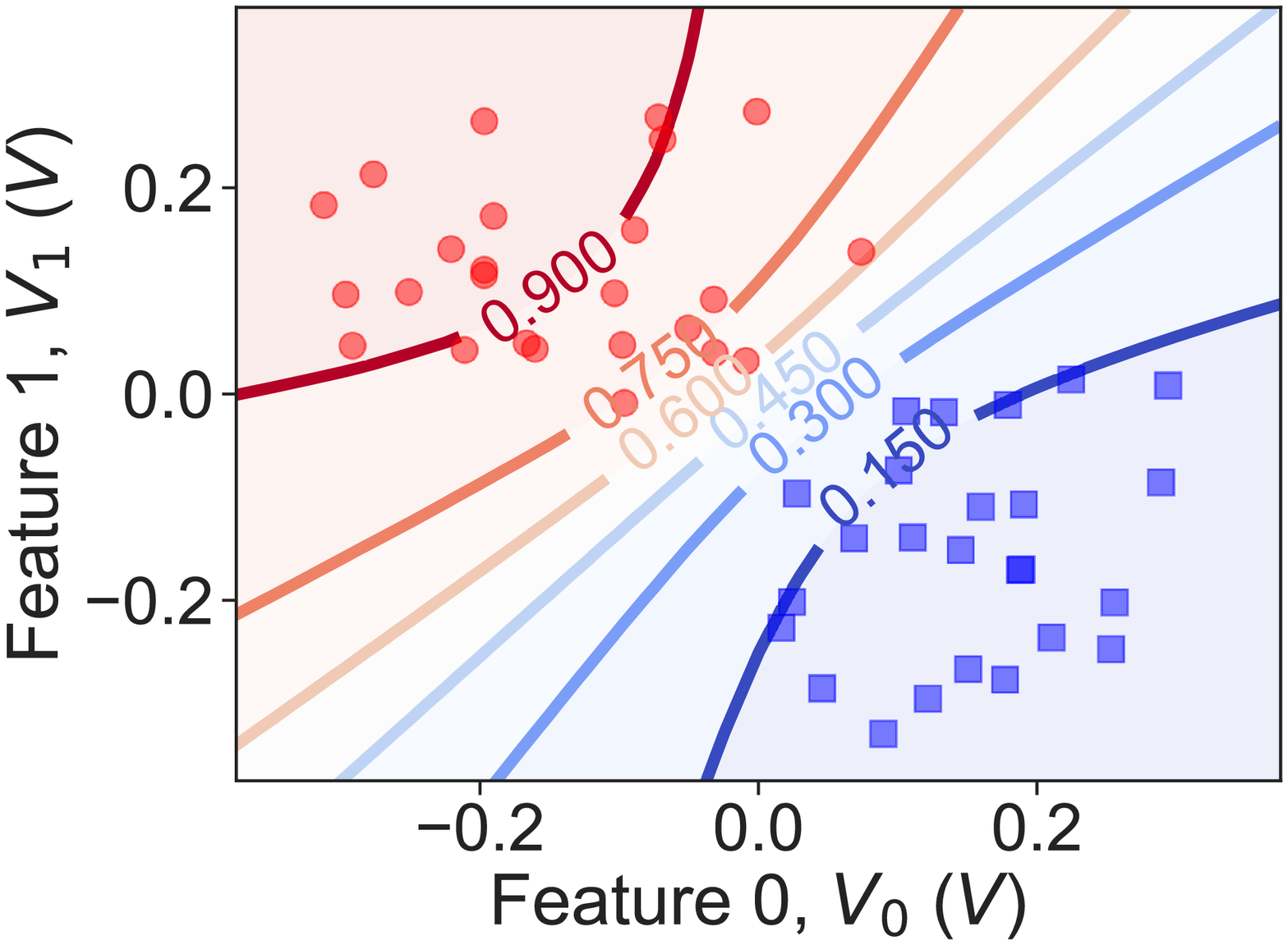}%
    \caption{}\label{fig3:probabilitycontour}
    \hfill
  \end{subfigure}
  \caption{
  \textbf{Experimental results on the illustrative 2-D dataset.}
  (\subref{fig3:randomwalk}) Posterior distribution stored within the memory array after the training experiment. The two conductance parameters of each accepted model are plotted as points in the conductance plane (or model space). The initial model stored in the zeroth row is shown by a red dot. The models accepted into the subsequent array rows are plotted as green points with an opacity proportional to the associated row counter value. The transparent lines between green points show the jumps on the posterior made between successive array rows. The resulting posterior distribution is superimposed in a contour plot whereby blue and green contours denote low and high probability density respectively. 
  (\subref{fig3:twoclasshyperplanes}) 
  The two classes of data (red circles and blue squares) and a subset of fifteen models stored in randomly selected rows of the memory array.
  %Each data point, sampled from one of two offset normal distributions, is plotted in the voltage encoded feature-space. Fifteen dashed black lines are superimposed and correspond the fifteen randomly selected models from the posterior.
  (\subref{fig3:probabilitycontour}) The probabilistic boundary that is described by the posterior distribution stored within the resistive memory array. Each of the contour lines is annotated with a probability that corresponds to the probability that any point lying on it belongs to the class of the red data. The bounded regions between contours are coloured from red to blue whereby red denotes high confidence that a point within that shaded region belongs to the red class, and blue a low confidence.
  }
\end{figure}
%TC:endignore

%\clearpage
%TC:ignore
\begin{figure}[!hp]
  \begin{subfigure}{0.36\linewidth}
    \centering
    \includegraphics[width=1\linewidth]{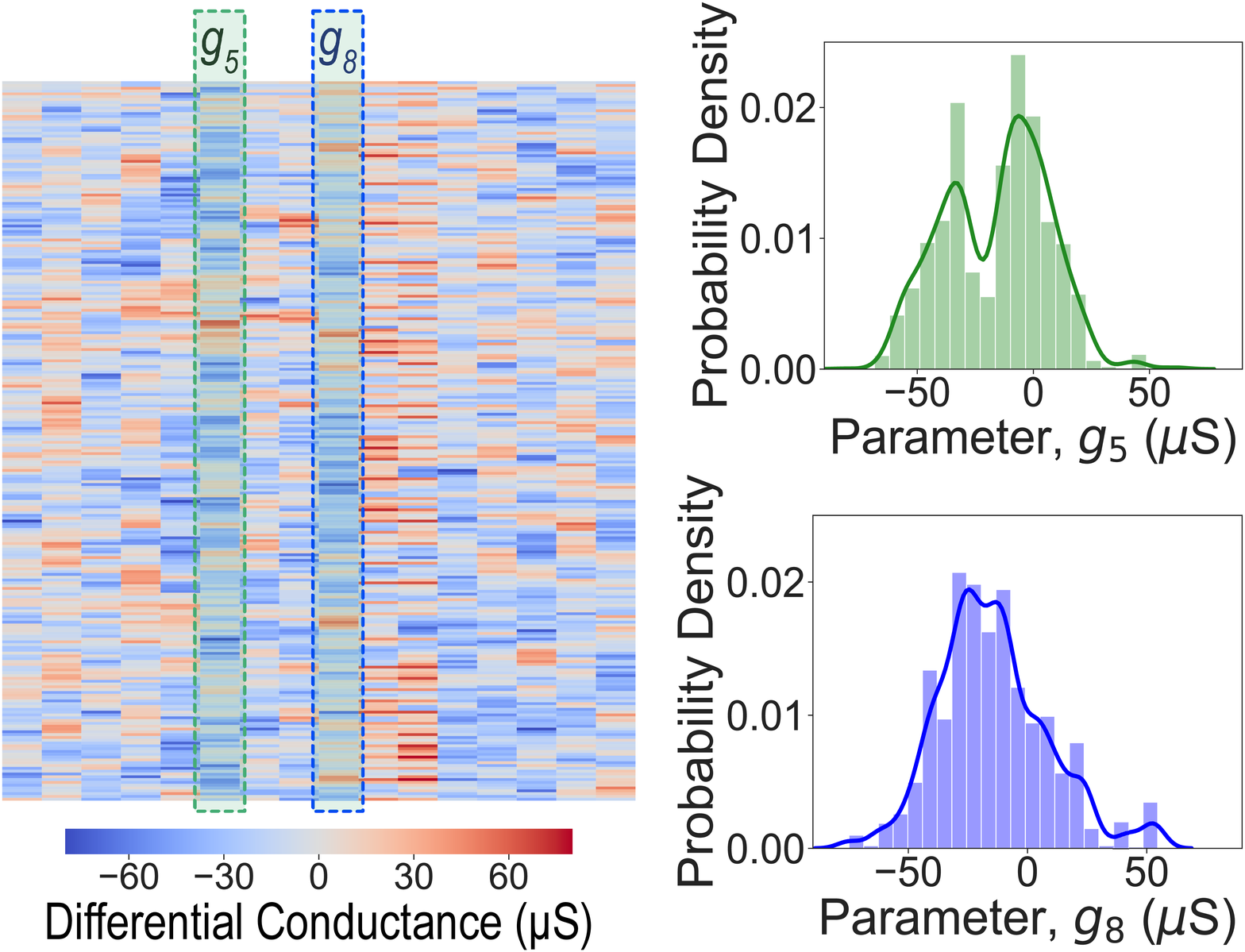}%
    \caption{}\label{fig3:trainedarray}
    \hfill
  \end{subfigure}
  ~
  \begin{subfigure}{0.30\linewidth}
    \centering
    \includegraphics[width=1\linewidth]{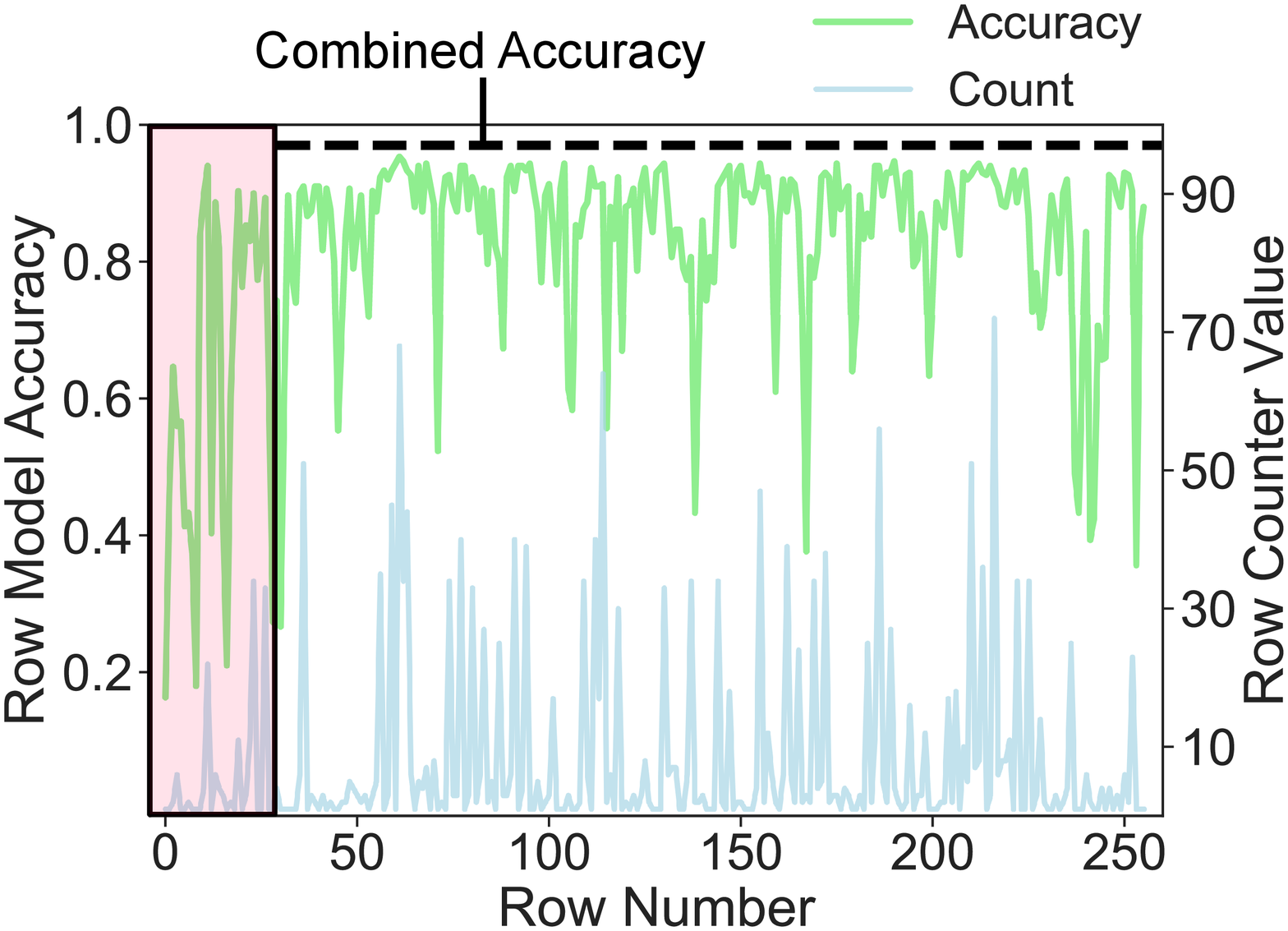}%
    \caption{}\label{fig3:accuracycount}
    \hfill
  \end{subfigure}
  ~
  \begin{subfigure}{0.30\linewidth}
    \centering
    \includegraphics[width=1\linewidth]{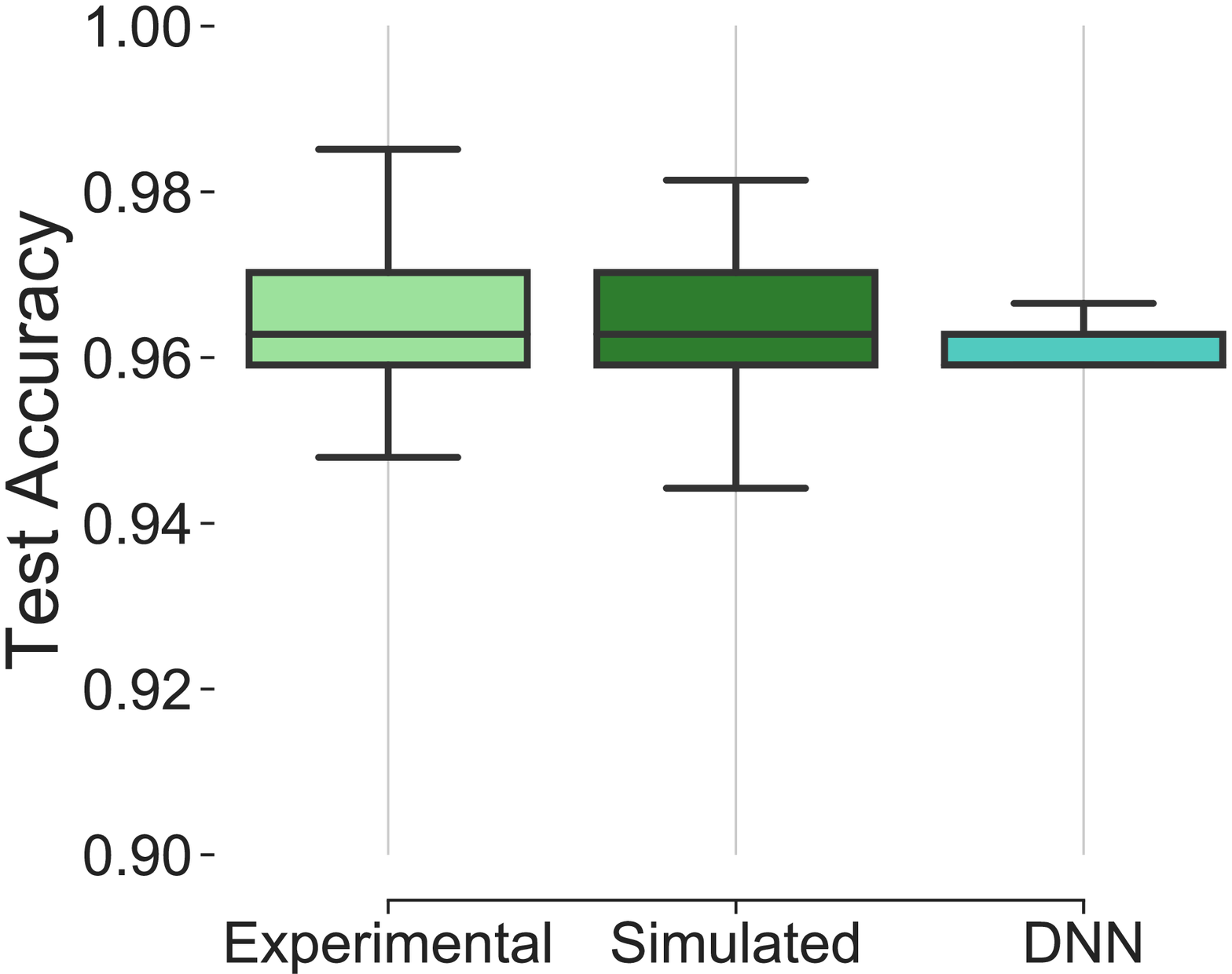}%
    \caption{}\label{fig3:supervisedaccuracy}
    \hfill
  \end{subfigure}
  ~
  \begin{subfigure}{0.30\linewidth}
    \centering
    \includegraphics[width=0.5\linewidth]{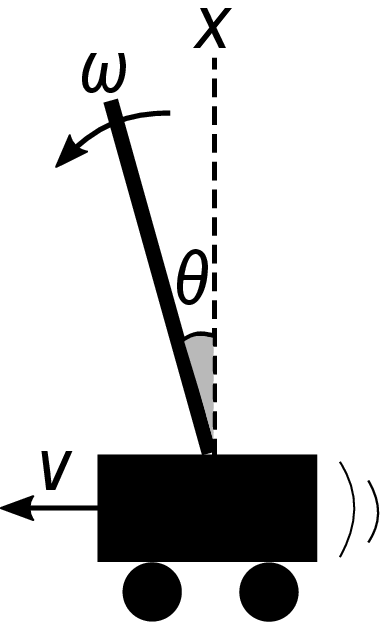}%
    \caption{}\label{fig3:cartpoledrawing}
    \hfill
  \end{subfigure}
  ~
  \begin{subfigure}{0.30\linewidth}
    \centering
    \includegraphics[width=1\linewidth]{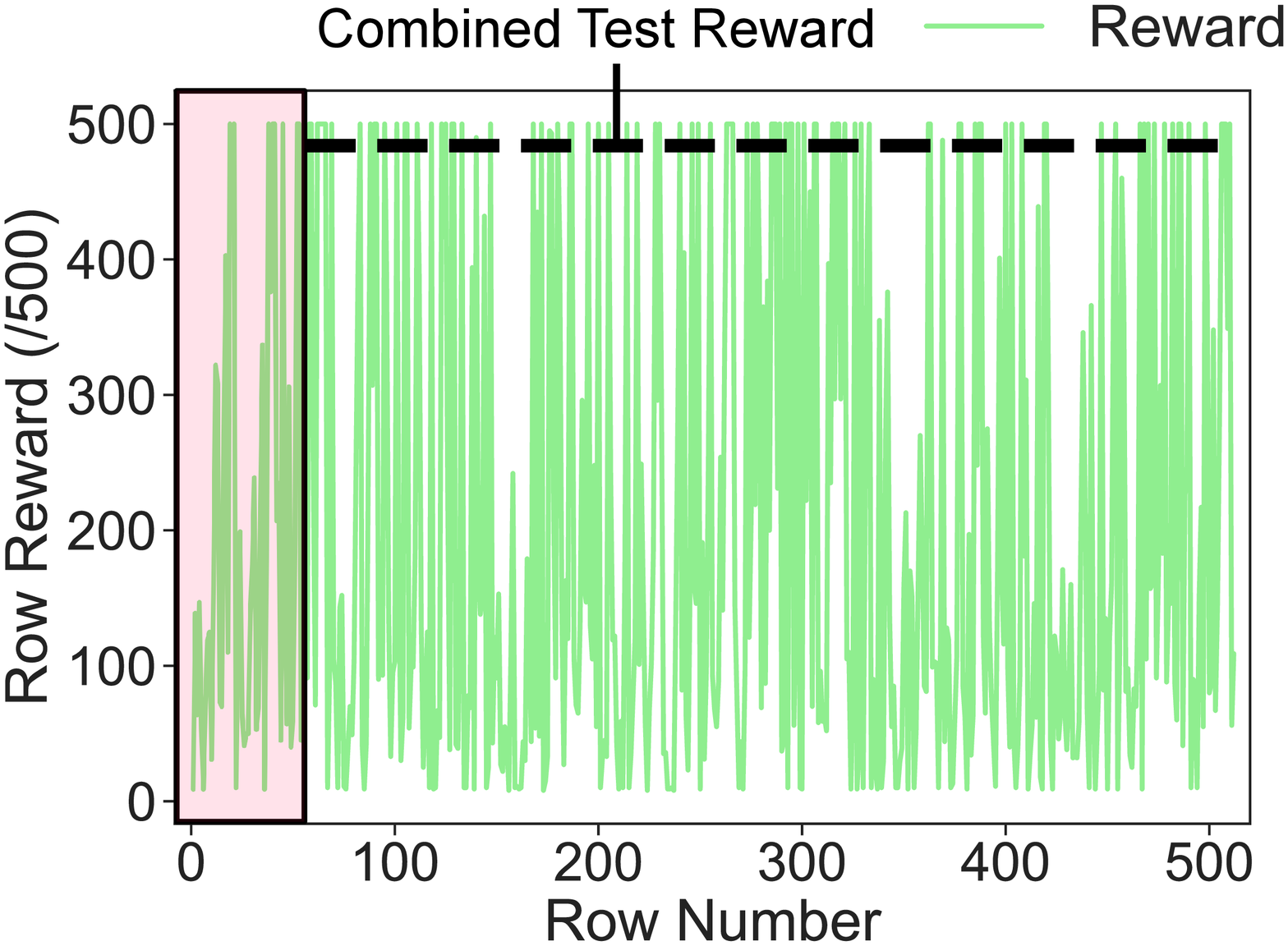}%
    \caption{}\label{fig3:rewardcount}
    \hfill
  \end{subfigure}
  ~
  \begin{subfigure}{0.30\linewidth}
    \centering
    \includegraphics[width=1\linewidth]{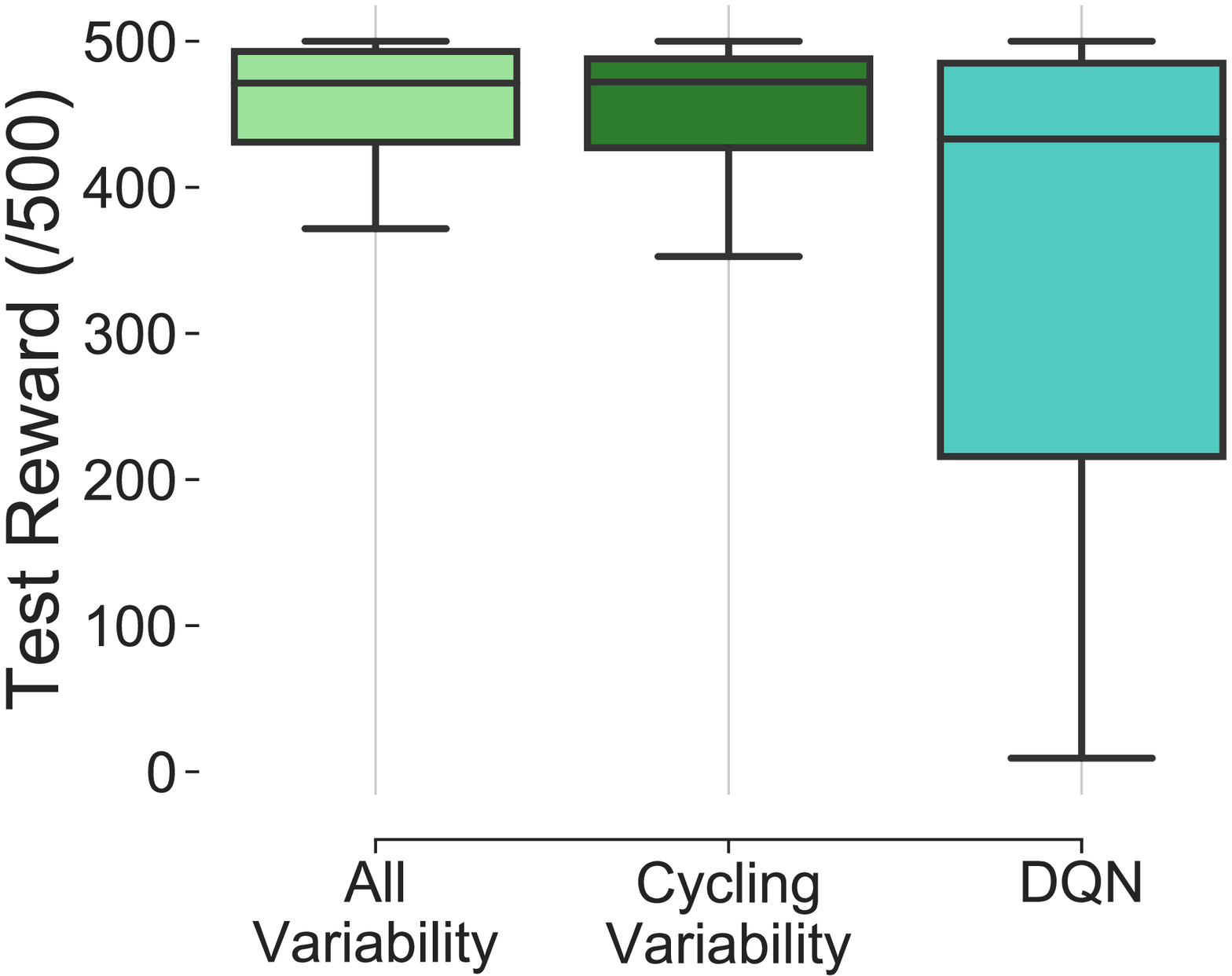}%
    \caption{}\label{fig3:reinforcementreward}
    \hfill
  \end{subfigure}
  ~
%\end{figure}
%\begin{figure}[p]
%\ContinuedFloat
\caption{
  \textbf{Experimental results on the supervised classification of breast tissue samples.} 
  (\subref{fig3:trainedarray}) (left) Heatmap of the differential conductance pairs in the $256\times16$ array after a training experiment. Cells within the heatmap are coloured from blue to red indicating the sign and magnitude of each conductance parameter. The row counter weighted distributions within columns five and eight, corresponding to two learned model parameters, are plotted in respective green and blue histograms and fitted with kernel density estimations.
  (\subref{fig3:accuracycount}) Accuracy, evaluated on the dataset, of the conductance model (green) and the row counter value (blue) for each of the $256$ rows. The first 32 rows, contained within a red rectangle, have been accepted during the burn-in period. Discarding these rows, the combined accuracy of the remaining rows on the test dataset is 97\%, indicated by the horizontal dashed line. 
  (\subref{fig3:supervisedaccuracy}) Boxplots showing the test accuracy distributions over 100 separate train/test iterations using the same train/test split for (left, light green) the experimental setup, (centre, dark green) the behavioural simulation and (right, blue) a neural network benchmark model. The coloured boxes span the upper and lower quartiles of accuracy while the upper and lower whiskers extend to the maximum and minimum accuracies obtained over the $100$ iterations. The median accuracy is indicated with a solid horizontal line.
  \textbf{Behavioural simulation results on the Cartpole reinforcement learning task.}
  (\subref{fig3:cartpoledrawing}) Diagram of the Cartpole task: learning how to accelerate left or right in order to maintain a pole balanced on top of a pole within 15 degrees of normal (vertical dashed line). The environment is described by four features; $X$ - x-position of the cart, $\theta$ - angle of the pole to vertical, $\omega$ - angular velocity at the tip of the pole and $\nu$ - velocity of the cart.
  (\subref{fig3:rewardcount}) Reward obtained during the training episode when each of the conductance models was accepted into each row of the $512$ memory array rows (the maximum reward is $500$). A burn-in period of $64$ rows is denoted with the red rectangle. The mean reward obtained over $100$ test episodes using the combination of the models accepted after the burn-in, equal to $484$ out of a maximum score of $500$, is denoted with a horizontal dashed black line.
  (\subref{fig3:reinforcementreward}) Boxplots showing the distribution of the mean test reward obtained during $100$ testing episodes achieved in the Cartpole task over 100 separate train/test iterations. (left, light green) Behavioural simulation considering device-to-device and cycle-to-cycle variability, (centre, dark green) behavioural simulation considering only cycle-to-cycle variability and (right, blue) a DQN benchmark model. The coloured boxes span the upper and lower quartiles of mean reward, the upper and lower whiskers extend to the maximum and minimum mean rewards obtained, and the median mean reward is indicated with a solid horizontal line.
}
\end{figure}
%TC:endignore

\section*{Supervised Learning}

We now demonstrate how this RRAM-based Metropolis-Hasting MCMC sampling algorithm can be used to address supervised learning tasks. This is achieved experimentally using the fabricated OxRAM array pictured in Fig.~\ref{fig2:arrayimage} with a computer-in-the-loop. The computer configures voltage waveforms, which iteratively read and program the devices in the array. It also calculates the acceptance ratio which then determines the subsequent programming operations that are applied (see Methods and supplementary Fig.~1).
%As a final demonstration we apply the same approach to reinforcement learning through a behavioural simulation.
%For each the likelihood expression used in Eq.~\ref{eq:acceptanceratio} is adapted.
%To apply resistive memory based MCMC sampling to a supervised learning task,
%we use the logistic function to operate on the current flowing out of each array row (see Methods).
% Fig.~\ref{fig2:bayesianmodel} ($f(\textbf{V}\dot~\textbf{g})=1/(1+exp(-S(\textbf{V}\dot~\textbf{g})))$). 
%This configures the array as a Bayesian logistic regression model\cite{Green95} which can be trained to output the probability that a previously unseen input data point, $\textbf{V}\textsubscript{new}$, belongs to one, of potentially many, classes.

First, as an illustrative example, we train a $2048\times2$ array to separate two classes of artificially generated data - the red circles from blue squares in Fig.~\ref{fig3:twoclasshyperplanes} (see Methods).
% , a linearly separable two-class data set is generated (red circles and blue squares in Fig.~\ref{fig3:twoclasshyperplanes}). Experimentally, using the computer-in-the-loop with a fabricated $2048\times2$ array of OxRAM devices (see Methods), we perform RRAM-based MCMC sampling with the objective of learning the posterior approximation that allows the array to recognise the red class of data points.
After the algorithm terminates, 
%at the $2047\textsuperscript{th}$ row, 
the 
%ensemble of OxRAM conductances and row counter values in the array give
non-volatile conductance states of the devices in the array give
rise to the multi-modal posterior approximation plotted in Fig.~\ref{fig3:randomwalk}. 
%From an initial conductance model (denoted with a large red circle), the algorithm has proceeded to propose new conductance models in subsequent array rows and thereafter accepted a further $2047$ counter weighted conductance models (small green circles with an opacity proportional to its row counter). 
Two distinct peaks emerge, denoting regions of high probability density where many of the accepted models
%, with high row counter values, 
are tightly packed. %into two regions on the plane. 
A randomly selected subset 
%of fifteen 
of these accepted models are plotted as hyper-planes in the space of the data in Fig.~\ref{fig3:twoclasshyperplanes}, 
each defining a unique linear boundary separating the two clouds of data points belonging to each class. 
By combining of all of the accepted models, therein using the posterior approximation which now exists in the array, a probabilistic boundary between the two classes emerges in Fig.~\ref{fig3:probabilitycontour}. 
%In contrast to a deterministic logistic regression model where parameters are single conductance values (Fig.~\ref{fig5:dotprod}), this Bayesian variant of the model describes probability contours that bend around and encapsulate the data points within regions of predictive confidence. 
Any 
%new, 
previously unseen data point can hereafter be assigned a probability of belonging to 
the class of red circles
% a class
as a function of where it falls on this probability contour.

We next apply the experimental system to a more realistic task, 
%in 
the classification of 
%$569$ 
histologically stained breast tissue as malignant or benign\cite{Wolberg91}, using a $256\times16$ array (see Methods). 
%is trained to recognise malignant tissue samples.
%using a training subset of 300 randomly selected points. 
After training has been completed, the 
%differential conductance 
parameters %which have been left imprinted 
programmed
into the resistive memory array are plotted in a heatmap in Fig.~\ref{fig3:trainedarray}. 
%The probability distributions of the differential conductance pairs in two columns of the memory array, therein two learned parameters of the Bayesian model, are also plotted alongside. 
The distributions of two of the learned parameters from the resulting posterior are shown alongside. 
%These distributions encapsulate the degree of uncertainty in the estimation of each parameter, for example the bi-model probability distribution that has been captured for the fifth parameter plotted in green.
In order to visualise the learning process, the classification accuracy of the accepted conductance model in each array row, in addition to its corresponding counter value, are plotted in green and blue traces respectively in Fig.~\ref{fig3:accuracycount}. From an initial conductance model, which achieves a poor classification accuracy, the algorithm quickly converges onto the posterior after approximately $32$ rows %during what is termed the 
of burn-in.
After this burn-in period, the algorithm tends to accept models into array rows which have a higher probability density, corresponding to models with higher classification accuracy. The rows containing higher accuracy conductance models also tend to have higher associated row counter values. Strikingly, the accuracy does not saturate but rather increases sharply during burn-in and then proceeds to oscillate between high and medium accuracy conductance models. This is an important property of MCMC sampling algorithms whereby sub-optimal models are also accepted, although less frequently and with a smaller row counter value, ensuring that the true form of probability density of the posterior is uncovered. 
After the algorithm terminates, the accuracy achieved 
on the testing set
was $97\%$, indicated by the horizontal dashed black line in Fig.~\ref{fig3:accuracycount}. 
Notably, the combined accuracy of all of the models in the posterior approximation is greater than the accuracy of any of the single accepted deterministic models alone. 

In order to gather statistics on the variability between training iterations, the training process was repeated in $100$ further experiments and the resulting accuracy distribution is reported in
% median accuracy over $100$ separate training iterations, using the same memory array, 
%the light green boxplot of 
Fig.~\ref{fig3:supervisedaccuracy}, achieving a median accuracy of 
%evaluated and is - where the median classification accuracy of malignant data points was 
$96.3\%$. In order to benchmark this result, a software-based 
%neural network 
neural network
model employing a single hidden layer and a 
%of thirty-four rectifying non-linear units (ReLus), such that the number of synapses in the neural network, which are themselves 32-bit floating point elements, 
total number of synaptic elements equal to the number of differential conductance pairs in the resistive memory array was applied to the same task (see Methods). The resulting median accuracy of the neural network benchmark was $95.8\%$ (Fig.~\ref{fig3:supervisedaccuracy}). The advantage of the RRAM-based MCMC sampling approach is 
% intensified
% compounded 
strengthened
by the fact that, while in the experimental array each parameter is described by the differential conductance between a pair of non-volatile nanoscale devices, 
each synaptic weight in the 
neural network
model is a 32-bit floating point precision 
number.

Calibrated on array level measurements of cycle-to-cycle and device-to-device variability, a behavioural simulation of the system was also developed (see Methods). The simulator was 
verified by applying it to the same supervised learning task as the experimental system. 
As seen in
Fig.~\ref{fig3:supervisedaccuracy}, %where it is seen to 
the results of the simulation
match closely
the results obtained 
experimentally, and can thus
serve as a 
tool in determining how the approach can be applied to further tasks.

\section*{Reinforcement Learning}

Finally, we demonstrate that the approach can 
be applied to reinforcement learning tasks using the developed behavioural simulator. In contrast to supervised learning, reinforcement learning does not require a labelled dataset that a model must learn to classify. Instead, a model is tasked with determining the actions of an agent in a physical or simulated environment in real-time \cite{Sutton98}. The agent observes, at each timestep, input information regarding the current state of the environment ($\textbf{V}$) and, as a function of the actions taken by the agent ($\textbf{a}$), a scalar reward ($r$) is received. 
The objective in reinforcement learning is to realise a model (often referred to as a policy) that allows an agent to take actions in an environment which maximise its expected reward. 
Here we apply RRAM-based MCMC sampling as a policy-search algorithm \cite{Hoffman07} and learn a posterior distribution in terms of reward.

Specifically we address the Cartpole control task \cite{Barto83} (see supplementary videos 1 and 2): an agent must learn how to control a pole balanced on top of a cart by accelerating to the left or right as a function of four observed environmental variables describing the velocity and position of the cart and pole (Fig.~\ref{fig3:cartpoledrawing}). To achieve this, we propose to use two $512\times4$ memory arrays (identical to that in Fig.~\ref{fig2:bayesianmodel}); one array must learn when to accelerate left and the other when to accelerate right. The rows of equivalent index, containing the proposed models in both arrays, are programmed together at the beginning of each training episode and are used during the episode to determine the actions taken by the agent in a winner-take-all fashion at each timestep. After the algorithm arrives to the final rows, the training period terminates and, as in the in supervised case, all of the accepted models are combined to determine the actions of the agent during testing episodes (see Methods).

To visualise the training process, the reward received by the agent for each accepted pair of models is plotted in Fig.~\ref{fig3:rewardcount} (for corresponding row counter values see supplementary Fig.~2). The reward is seen to oscillate during training as the algorithm explores the posterior distribution. After training, the agent then uses the learned posterior approximation to select actions over $100$ testing episodes - achieving a mean test reward of $484$ out of $500$ (see Methods). To determine the variability between training iterations, this procedure was repeated $100$ times and the distribution of mean test reward is plotted in Fig.~\ref{fig3:reinforcementreward}. Over the $100$ training iterations the median mean test reward obtained was $475$. In order to benchmark this result, a Deep-Q Network \cite{Mnih15} (DQN) reinforcement learning model was applied to the same task (see Methods). The DQN employed one hidden layer of neurons with a total number of synapses equal to the number of differential conductance pairs used in the two memory arrays. The distribution of mean test reward obtained by the DQN is plotted in Fig.~\ref{fig3:reinforcementreward} where it obtained a median mean test reward of only $420$ - less than that of the resistive memory array - while also exhibiting greater variability between training iterations.

In order to assess the impact of device-to-device variability (Fig.~\ref{fig1:hcsdevicetodevice}), which traditionally negatively impacts the gradient-based training of RRAM-based neural network models, the simulation was repeated without its consideration (see Methods). The resulting mean test reward distribution is plotted in Fig.~\ref{fig3:reinforcementreward} where it is seen to be largely equivalent to that case where device-to-device variability was considered.
% - albeit with a slightly reduced median test reward of $471$. 
This result challenges the longstanding conception that device-to-device variability is a disadvantage in RRAM-based machine learning and requires mitigation. However, this should not necessarily come as a surprise: unlike in gradient-based optimisation, where device-to-device variations impede the proper descent down an error gradient, MCMC sampling centres on the random generation of models. From this perspective the device-to-device variability that is incurred when moving between rows of the memory array (Fig.~\ref{fig1:hcsdevicetodevice}) can be viewed as another means of local exploration on the posterior - an additional computational mechanism.

\FloatBarrier

\section*{Conclusion}

Resistive memories promise to be the key in realising energy-efficient intelligent computing systems that can respond and learn locally at the edge due to their capacity to compute in-memory. In this paper, we realised that the potential of resistive memories to be employed as physical random variables can be combined with their established high efficiency dot-product capabilities to permit the implementation of in-memory Markov Chain Monte Carlo sampling algorithms. In a significant departure from previous work where gradient-based algorithms have struggled to extract the true potential from RRAM, due to the mismatch between device and algorithm, we have found RRAM to be an ideal substrate for MCMC sampling. Furthermore, the presence of device-to-device variability was not found to be an impediment to performance; indicating that area, energy or time intensive mitigation techniques, essential in gradient-based approaches, are not required to realise a practical system.

Using a computer-in-the-loop experiment with a fabricated array of oxide-based filamentary resistive memories, we experimentally applied RRAM-based MCMC sampling to train, in-situ, the array, which is configured as a Bayesian machine learning model, to address two supervised learning tasks. We then showed that the same approach can also be applied to reinforcement learning using an accurate behavioural simulator. In each case RRAM-based MCMC sampling outperformed a software-based neural network model realised with an equivalent number of memory elements. 

We aim to move from this successful experimental demonstration by taking the computer out of the loop and
%co-integrating equivalent CMOS circuits with an RRAM array to
producing a standalone chip that can be applied to higher complexity tasks outside of the laboratory. 
%to discover use cases for our approach.
In order to scale to such tasks, just as a layers of interconnected logistic regression models realise a neural network model, we will explore how Bayesian networks and Bayesian neural networks can be constructed by networking together resistive memory arrays on-chip. 
Ultimately, we have shown that, by embracing what have been previously considered as non-ideal device properties, resistive memory based Markov Chain Monte Carlo sampling algorithms can sit at the core of a new generation of intelligent energy-constrained computing systems - capable of responding, adapting and learning locally at the edge.
%TC:ignore

\section*{Acknowledgements}
The authors would like acknowledge the support of the French ANR via Carnot funding as well as the European Research Council (grant NANOINFER, number 715872) for the provision of funding. In addition, we would like to thank E.~Esmanhotto, J.~Sandrini and C.~Cagli (CEA-Leti) for help with the measurement setup, J.F.~Nodin (CEA-Leti) for providing the images in Fig.~\ref{fig2:arrayimage} and to S.~Mitra (Stanford University), M.~Payvand (ETH Zurich), A.~Valentian, M.~Solinas-Angel, E.~Nowak (CEA-Leti), J.~Diard (CNRS, Universit\'e Grenoble Alpes), P.~Bessière and J.~Droulez (CNRS, Sorbonne Universit\'e) and J.~Grollier (CNRS, Thales) for discussing various aspects of the paper. 

\section*{Author Contributions}
TD developed the concept of RRAM-based MCMC sampling. NC built the computer-in-the-loop test setup with the resistive memory array. TD and NC performed the computer-in-the-loop experiments with the resistive memory array. TD implemented the behavioural simulator, performed measurements on the array which were used to calibrate the simulation and implemented the neural network benchmark models. TD, DQ and EV developed ideas for and wrote the paper together.

\section*{Conflict of Interest Statement}
The authors declare no conflict of interest.

\section*{Methods}

\subsection*{Hafnium Dioxide Based Resistive Memory Arrays and Experimental Setup}
Two versions of fabricated OxRAM memory arrays are used in the presentation of the paper. The first is a $4,096$ (4k) device array (16$\times$256 devices) of 1T1R structures. The second chip is a $16,384$ (16k) device array (128$\times$128 devices) of 1T1R structures. In each array, the OxRAM cell consists of a HfO\textsubscript{2} thin-film sandwiched in a TiN/HfO$_2$/Ti/TiN stack. The HfO\textsubscript{2} and Ti layers are 10~nm tick and have a mesa structure 300~nm in diameter. The OxRAM stack is integrated into the back-end-of-line of a commercial 130~nm CMOS process. In the 4k device array the n-type selector transistors are 6.7$\mu$m wide. In the 16k array the n-type selector transistors are 650nm wide. Voltage pulses, generated off chip, can be applied across specific source (SL), bit (BL) and word lines (WL) which contact the OxRAM top electrodes, transistor sources and transistor gates respectively. External control signals determine to what compliment of SL, BL and WL the voltage pulses are applied over by interfacing with CMOS circuits integrated with the arrays. Signals for the 4k device array are generated using the Keysight B1530 module and those for the 16k device array are generated by the RIFLE NplusT engineering test system. The RIFLE NplusT system can also run C++ programs that allow the system to act as a computer-in-the-loop with the 16k device array.

Before either chip can be used, it is required to form all the devices in the array. In the forming process oxygen vacancies are introduced into the HfO\textsubscript{2} thin-film through a voltage-induced dielectric breakdown. This is achieved by selecting devices in the array one at a time, in raster scan fashion, and applying a voltage between the source and bit lines. At the same time, the current is limited to the order of $\mu$As by simultaneously applying an appropriate V\textsubscript{WL} (transistor gate) voltage. A form operation consists of the following conditions;
4k device array - V\textsubscript{SL}=4V, V\textsubscript{BL}=0V, V\textsubscript{WL}=0.85V, 16k device array - V\textsubscript{SL}=4V, V\textsubscript{BL}=0V, V\textsubscript{WL}=1.3V. After the devices have been formed, they are conditioned by cycling each device in the array between the LCS and the HCS one hundred times. Unless otherwise specified, the standard RESET conditions used in the paper were; 4k device array - V\textsubscript{SL}=0V, V\textsubscript{BL}=2.5V, V\textsubscript{WL}=3V, 16k device array - V\textsubscript{SL}=4V, V\textsubscript{BL}=0V, V\textsubscript{WL}=2.5V. Unless otherwise specified, the standard SET conditions used were; 4k device array - V\textsubscript{SL}=2V, V\textsubscript{BL}=0V, V\textsubscript{WL}=1.2V, 16k device array - V\textsubscript{SL}=1.8V, V\textsubscript{BL}=0V, V\textsubscript{WL}=2.0V. The device conductances are determined by measuring the voltage drop over a known low-side shunt resistance connected in series with the selected SL in a read operation. 
%The current flowing out of the source line is given by dividing the measured voltage by the known resistance value. The conductance can then be calculated by dividing this measured current by the applied read voltage. 
Devices are read according to the following conditions; 4k device array - V\textsubscript{SL}=0.1V, V\textsubscript{BL}=0V, V\textsubscript{WL}=4.8V, 16k device array - V\textsubscript{SL}=0.4V, V\textsubscript{BL}=0V, V\textsubscript{WL}=4.0V. All off-chip generated voltage pulses for programming and reading have a pulse-width of 1$\mu$s. 

\subsection*{Measurement of OxRAM HCS Random Variable Properties}
To characterise properties of the HfO\textsubscript{2} physical random variable, as plotted in Figs.~\ref{fig1:hcscycletocycle}, \ref{fig1:hcspowerlaw} and \ref{fig1:hcsdevicetodevice}, the 4k device array chip was used. This allows use of a larger selector transistor which offers a greater range of the SET programming currents. 
%less transistor variability 
%and a smaller parasitic series resistance during a read operation than for the 16k device array. 
To measure the data plotted in Fig.~\ref{fig1:hcspowerlaw} a 4k device array was formed, conditioned and then RESET/SET cycled one hundred times over a range of nine word line voltages (V\textsubscript{WL}), corresponding to the range of SET programming currents in Fig.~\ref{fig1:hcspowerlaw}. Each device was read after each SET operation. Between each step in V\textsubscript{WL}, the devices were additionally RESET/SET cycled 100 times under standard programming conditions ensuring that, for each of the hundred cycles at different V\textsubscript{WL}, the initial conditions were the same. For the data plotted in Figs.~\ref{fig1:hcscycletocycle} and \ref{fig1:hcsdevicetodevice} a single device and all 4k devices in the 4k device array were respectively RESET/SET cycled 500 times. The conductance was read after each SET operation. This conductance data was then processed and plotted using the python libraries NumPy, SciPy, Seaborn and Matplotlib. 

\subsection*{MCMC Sampling Experiments on the 16k Device Array}
The computer-in-the-loop experiments (used to obtain the results in Figs.~\ref{fig3:randomwalk}, \ref{fig3:twoclasshyperplanes}, \ref{fig3:probabilitycontour}, \ref{fig3:trainedarray}, \ref{fig3:accuracycount} and \ref{fig3:supervisedaccuracy}) made use of the 16k device array interfaced to the C++ programmable RIFLE NplusT system (see supplementary Fig.~1). The devices in the array, which physically exist as a 128$\times$128 array of 1T1R structure, are re-mapped into a virtual address space which realises the structure presented in Fig.~\ref{fig2:bayesianmodel}. This is achieved by allocating pairs of sequential banks of $M$ devices (for an $M$-parameter model) corresponding to the $\textbf{g}\textsubscript{+}$ and $\textbf{g}\textsubscript{-}$ conductance vectors to each of the $N$ rows in Fig.~\ref{fig2:bayesianmodel}. A dot-product is performed by reading the conductances of the devices composing $\textbf{g}\textsubscript{+}$ and $\textbf{g}\textsubscript{-}$ and subtracting them in the computer-in-the-loop to arrive at $\textbf{g}$ and then performing the dot-product between $\textbf{V}$ and $\textbf{g}$. Note that in a future version of the system, by integrating appropriate circuits within the device array the dot-product can be performed by simply applying the data points as read voltages and reading the output current as described in the paper.

Resistive memory based MCMC sampling begins by performing a RESET operation on each device in the array, rendering all devices in the LCS. Using the variable $n$ to point to the row containing the current model, the algorithm begins with $n=0$. The devices in row $n$ are SET. Because we do not have strong prior beliefs on model parameters, each device is SET using the lowest available V\textsubscript{WL} (in this case $1.4V$) which corresponds to the lowest SET programming current (20$\mu$A). As a result the standard deviation of the initial samples will be high (Fig.~\ref{fig1:hcspowerlaw}), thereby capturing this uncertainty. The devices in the following row, $n+1$, are then programmed with SET programming currents proportional to the conductances read from the corresponding devices in row $n$. This results in a proposed model, $\textbf{g}\textsubscript{p}$ being generated in row $n+1$ inline with the proposal distribution offered by the cycle-to-cycle HCS conductance variability:

\begin{equation}\label{eq:setproposal}
p \left(\textbf{g}\textsubscript{p}|\textbf{g} \right)
=\mathcal{N} \left(\textbf{I}\textsubscript{SET}(\textbf{g}),\sigma(\textbf{g}) \right).
\end{equation}

In doing this, a new conductance value is sampled for each device in row $n+1$ from a normal random variable with a median value corresponding to the same device in row $n$, offset by device-to-device variability (Fig.~\ref{fig1:hcsdevicetodevice}) that is introduced when moving between successive rows.

The SET programming current in the proposal step is determined by the value of V\textsubscript{WL} used to program each device in row $n+1$. To achieve this a single
look-up table is determined in an initial sweep step whereby the entire 16k device array is RESET/SET cycled once per V\textsubscript{WL} value that will be used in the experiment. For each of these values of V\textsubscript{WL}, the median conductance read across the 16k device array is calculated and inserted in the corresponding entry in the table. Therefore, when programming a device in the row containing the proposed model is required to read the conductance of the corresponding device in row $n$, and use the value of V\textsubscript{WL} with the closest corresponding conductance, as the V\textsubscript{WL} used in the SET operation. In the experiments, the look-up table extended from 1.4V to 1.8V in discrete 20mV steps, corresponding to SET programming currents in the range of 20$\mu$A-100$\mu$A and permitted
% This exploited the range of V\textsubscript{WL} up until the effect of selector transistor saturation,
%the conductance response and
permitting median conductances in the range of 40$\mu$S- 80$\mu$S to be used (see supplementary Fig. 3).
Programming the devices in row $n+1$ implements the model proposal step depicted in Fig.~\ref{fig2:proposal}. 
%As a precaution against proposing models at rows with non-functioning devices, potentially causing the MCMC sampling algorithm to become stuck, 
%A limit of 32 was imposed on the maximum number of proposals per virtual row. 
%In the experiments reported in this paper this limit was set to thirty-two. 
%When the number of proposals exceeded this limit another pair of $M$ device conductance models were virtually re-mapped to the $n+1$\textsuperscript{th} row.

The Metropolis-Hastings MCMC sampling, after proposing a new model, requires to make a decision on whether to accept and record the proposed model $\textbf{g}\textsubscript{p}$ or reject and record, once again, the current model $\textbf{g}$. This decision is made based on the calculation of a quantity named the acceptance ratio $a$. Because the proposal density is normally distributed (Equation \ref{eq:setproposal}), and therefore symmetrical, it can be written as:

\begin{equation}\label{eq:acceptanceratio}
a=\frac{p(\textbf{g\textsubscript{p}})}{p(\textbf{g})} \frac{p(\textbf{t}|\textbf{g\textsubscript{p}},\textbf{V})}{p(\textbf{t}|\textbf{g},\textbf{V})}.
\end{equation}
This acceptance ratio is a number proportional to the product of the likelihood of a proposed model ($p(\textbf{t}|\textbf{g\textsubscript{p}},\textbf{V})$) and a prior on the proposed model ($p(\textbf{g\textsubscript{p}})$), divided by the product of the likelihood and prior of the current model. Given a dataset of $D$ data points where $A$ data points belong to the class the model is required to recognise ($t=1$) and $B$ other data points that the model should not recognise ($t=0$) the likelihood of a model is given by:

\begin{equation}\label{eq:supervisedlikli}
p\left(\textbf{t}|\textbf{g},\textbf{V}\right)=
\prod _{a=0}^{A}
%(\frac{1}{1+e\textsuperscript{-S(\textbf{V\textsubscript{a,t=1}\textbf{g}})}})
f\left( \textbf{V}\textsubscript{a,t=1}\cdot\textbf{g} \right)
\times
\prod_{b=0}^{B}
%(1-\frac{1}{1+e\textsuperscript{-S(\textbf{V\textsubscript{b,t=0}\textbf{g}})}})
\left( 1 - f\left( \textbf{V}\textsubscript{b,t=0}\cdot\textbf{g} \right) \right)
.
\end{equation}
The function $f(\textbf{V}\cdot~\textbf{g})$ depends on the specific formulation of the model. The prior of a model is given by:

\begin{equation}\label{eq:priorcalculation}
p \left(\textbf{g} \right)=
\frac{1}{\sigma\sqrt{2\pi}}
exp \left( - {\frac{\left(\textbf{g}-\textbf{$\mu$}\right)\textsuperscript{2}}{2\sigma\textsuperscript{2}}} \right),
\end{equation}
where the constant $\sigma$, corresponds to the prior belief that the posterior distribution is a multi-dimensional normal distribution with a standard deviation of $\sigma$ in each dimension. 
In all examples in this paper the value of $\mu$ was set to zero. %such that the prior is normally distributed around zero. 
These quantities are calculated on the computer-in-the-loop in logarithmic scale.
%in order to avoid numerical under-flow resulting from the repeated multiplication of small numbers.
In order to decide if the proposed model should be accepted or rejected, $a$ is compared to a uniform random number between $0$ and $1$, $u$, generated on the computer using the C++ standard random math package. 
% Note that uniform number generation can also achieved using an additional row of OxRAM devices, exploiting the sub-threshold stochastic SET operation (see supplementary Figure X).
If $a$ is less than $u$ then $\textbf{g\textsubscript{p}}$ is rejected. This is achieved by programming the devices in row $n+1$ back into the LCS and incrementing the counter at row $n$, $C\textsubscript{n}$, by one. The counter is a variable in the C++ on the computer although, as is the case for all functionality of the computer-in-the-loop, will be integrated as a 
%digital
circuit on future implementations of the system. The devices in row $n+1$ are 
then SET once more under the same programming conditions - generating a new proposed model $\textbf{g\textsubscript{p}}$ at row $n+1$. This process repeats until $a$ is found to be greater than $u$. When this is the case, $\textbf{g\textsubscript{p}}$ is accepted whereby the counter at row $n+1$ is incremented by one and the model at row $n+1$ becomes the new current model ($n=n+1$). 
This new current model $\textbf{g}$ is then used to propose a model, $\textbf{g\textsubscript{p}}$, at the next row in the array. The model stored at row $n-1$ is then left preserved in the non-volatile conductance states of the resistive memory devices, weighted by the counter value $C\textsubscript{n-1}$. 
%If the number of proposals exceeds the proposal limit of $32$, the counter pointing to row $n$ is not reset to zero but continues to be incremented by one for each rejected proposal at the new, re-mapped, row.
As this process repeats, and progresses down the rows of the memory array, the algorithm randomly walks around the posterior distribution leaving information on its probability density imprinted into the non-volatile conductance states of the OxRAM devices and the row counter values. Upon arriving at the final row of the array, $n=N-1$, the training process terminates resulting in a physical array of resistive memory devices which contains an approximation of the posterior distribution that can then be used in inference. 

Performing inference consists of computing the dot-product between a new, previously unseen data point ($\textbf{V\textsubscript{new}}$), and the model recorded in each array row. The response of each row is then multiplied by the value in each row counter. The summed response of each row in the array is then divided by the sum of all row counter values. This results in a scalar value which has been inferred from the $N$ weighted samples from the posterior distribution approximation:

\begin{equation}\label{eq:supervisedinference}
P\left(T\textsubscript{new}=1|\textbf{V},\textbf{t}\right)
=
\frac{1}{Tot}
\sum_{n=\beta}^{N-1} 
%\frac{Count\textsubscript{n}}
%{1+e\textbf{\textsuperscript{-S(\textbf{V\textsubscript{new}\textbf{g}\textsubscript{n}})}}}
C\textsubscript{n} f \left( \textbf{V}\textsubscript{new}\cdot\textbf{g}\textsubscript{n} \right)
.
\end{equation}
The summation considers only rows of an index greater than $\beta$ which determines the number of rows discarded to account for the burn-in period. The variable $Tot$ is the sum of all row counter values recorded after the burn-in period. 

\subsection*{Supervised Learning Experiment}
The memory array used in the supervised learning experiments is configured as a Bayesian logistic regression model by using the row function block:

\begin{equation}\label{eq:supervisedfunction}
  f(\textbf{V}\cdot~\textbf{g})=\frac{1}{1+e\textsuperscript{-S(\textbf{V}$\cdot$\textbf{g})}},
\end{equation}
where $S$ is a scaling parameter.
This logistic function limits the response of the row dot-product into a probability between $0$ and $1$. %This function incorporates an additional scaling parameter $S$. 
%Without an appropriate scaling factor, the response of the dot-product evaluates as a small number due to the multiplication of a data point in the $mV$ regime and device conductances on the order of $\mu$S. The output of the logistic function would therefore be restricted to values very close to $0.5$. This becomes taskatic when computing the acceptance ratio (Equation.~\ref{eq:acceptanceratio}), as likelihoods for highly accurate and highly inaccurate models will have similar values. 
When configured as such the likelihood of a conductance model is calculated as:

\begin{equation}\label{eq:supervisedlikli_methods}
p\left(\textbf{t}|\textbf{g},\textbf{V}\right)=
\prod _{a=0}^{A}
\left(\frac{1}{1+e\textsuperscript{-S(\textbf{V\textsubscript{a,t=1}$\cdot$\textbf{g}})}}\right)
\times
\prod_{b=0}^{B}
\left(1-\frac{1}{1+e\textsuperscript{-S(\textbf{V\textsubscript{b,t=0}$\cdot$\textbf{g}})}}\right).
\end{equation}
After training, inference can performed on a new data-point $\textbf{V}\textsubscript{new}$ whereby it is assigned a probability of belonging to the class $t=1$ by computing:

\begin{equation}\label{eq:supervisedinference_methods}
P(T\textsubscript{new}=1|\textbf{V},\textbf{t})=\frac{1}{Tot}
\sum_{n=\beta}^{N-1} \frac{C\textsubscript{n}}{1+e\textbf{\textsuperscript{-S(\textbf{V\textsubscript{new}$\cdot$\textbf{g}\textsubscript{n}})}}}.
\end{equation}

In the first supervised learning task, the random sampling library from NumPy was used to generate $50$ samples from a 2-D normal distribution centred at origin. Half of the points were assigned a class label of $t=1$ and shifted up and to the left while the other half were shifted down and to the right (by the same value) and labelled $t=0$ - providing an artificial linearly separable dataset of two classes for an illustrative demonstration of the system.
For the second supervised learning task, the Wisconsin breast cancer dataset was used which consists of 569 data points with class labels malignant ($t=1)$ or benign ($t=0$) \cite{Wolberg91}. The dataset was accessed through the Scikit-learn library and the random shuffle function from NumPy was used to sort the dataset into $369$ training points and $200$ test points that were used in each of the $100$ train/test iterations. Using the implementation of the Chi2 feature selection algorithm \cite{Liu95} available through Scikit-learn the number of features was reduced to sixteen, as this corresponded to the number of array columns used in the experiment. A further data pre-processing step was performed using the scale function from Scikit-learn to centre the dataset around the origin such that the model does not require an additional bias parameter. If this step were not performed an extra column could be added to the array which would then learn the distribution of the bias parameter. During training, the algorithm was configured to recognise data points corresponding to malignant tissue samples ($t=1$). During inference time the $200$ previously unseen data-points from the test split were assigned a probability of being malignant using Equation~\ref{eq:supervisedinference_methods}. Output probabilities greater than or equal to $0.5$ corresponded to a prediction of the sample being malignant ($t=1$) and probabilities less than $0.5$ corresponded to a prediction of the sample being benign ($t=0$). The reported test accuracy corresponds to the fraction of the $200$ test data points which were correctly classified.

\subsection*{MCMC Sampling Behavioural Simulator}
A custom behavioural simulation of our experiment was developed in python implementing the presented Metropolis-Hastings Markov Chain Monte Carlo sampling algorithm. The proposal distributions were calibrated on the data measured on the 4k device array. A normal distribution, using the random function suite from the NumPy library, was used to sample proposed models ($\textbf{g}\textsubscript{p}$) from current models ($\textbf{g}$). The standard deviation of the sample was determined based on the data plotted in Fig.~\ref{fig1:hcspowerlaw}, where the median relationship was seen to follow the power-law:

\begin{equation}\label{eq:sdpowerlaw}
SD=a\times~I\textsubscript{SET}\textsuperscript{b},
\end{equation}
with constants $a=0.093~A^{1-b}$ and $b=0.48$.
%The fitting was performed using the curve fit function from the optimisation library in SciPy. 
The current $I\textsubscript{SET}$ is determined based on the conductances from the current model using the data in Fig.~\ref{fig1:hcspowerlaw} where:

\begin{equation}\label{eq:gpowerlaw}
%I\textsubscript{SET}=\sqrt[c]{\frac{g}{d}},
I\textsubscript{SET}=\left({\frac{g}{d}}\right) ^{1/c},
\end{equation}
%with the constants $c=0.78$ and $d=0.19$ 
with the constants $c=0.78$ and $d=0.19~S/A^{c}$ 
(plotted in Fig.~\ref{fig1:hcspowerlaw}). To incorporate device-to-device variability the standard deviation in $c$, fit for individual devices in the 4k device array (see supplementary Fig.~4), is and found equal to be $e=0.096~S/A^{c}$. 
%When moving between sequential rows during the simulation, the value of $g$ was offset to a new random value which serves as the median of future samples each device in the simulation:
%\begin{equation}\label{eq:gdevicetodevice}
%g\textsubscript{offset}=\sqrt[c]{\frac{g}{d}}^{\mathcal{N}(c,e)}.
%g\textsubscript{offset}=\left( \frac{g}{d} \right) ^{\mathcal{N}(c,e)/c}.
%\end{equation}
The likelihood and acceptance ratio calculations were performed in the log-domain.% After calculating $log(a)$ it was then compared to a uniform random number, generated using the uniform random generator from NumPy, $log(u)$.

\subsection*{Reinforcement Learning Simulation}
The python library gym was used to simulate the Cartpole environment. The Cartpole environment provides four features to the behavioural simulator at each timestep of the simulation. The behavioural simulator then specifies the actions to be taken by the agent in the environment at the next simulation timestep.
Two simulated $512\times4$ arrays were used. One array encoded the accelerate left action while the other encoded the accelerate right action. The two arrays compete in a winner-take-all fashion to determine the actions taken by the agent. During training the output of the array rows containing the two proposed models was used to determine the actions of the agent at each timestep. Under application of a new observation from the environment $\textbf{V}$ the response:
\begin{equation}\label{eq:reinforcementfunction}
  f(\textbf{V}\cdot\textbf{g})=S \;
  \textbf{V}\cdot\textbf{g},
\end{equation}
was calculated, where $S$ is a scalar constant. 
%Despite the fact that there are two memory arrays, 
The two arrays are treated as if they were a single model. Rows of equivalent index in both arrays share a common row counter and are programmed and evaluated at the same time. 
The devices within the zeroth row of both arrays are initially SET by sampling from a normal random variable with the lowest available conductance median (in this simulation the conductance range extended from 50$\mu$S-200$\mu$S). 
The initial model is evaluated during a training episode where the cumulative reward received during the episode is recorded. For each timestep that the agent does not allow the pole to rotate 15 degrees outwith of perpendicular, or does not move outwith the bounds of the environment (both resulting in early termination of the episode), the agent receives a +1 reward. If the agent maintains the pole balanced for $500$ timesteps the episode terminates resulting in an episode in which the agent has received the maximum possible score of $500$. Respective models are then generated at the first row of each array by sampling new models from normal random variables with medians equal to the conductances of the corresponding devices in the zeroth row, offset by device-to-device variability.
%(Equation~\ref{eq:gdevicetodevice}). 
The two proposed models determine the actions of the agent during the following training episode whereby the agent accelerates to either the left or the right at each timestep of the episode as a function of which array exhibited the greater response (Equation \ref{eq:reinforcementfunction}). As a function of the cumulative reward achieved during this training episode a decision is made on whether to accept or reject the proposed model using the acceptance ratio:
\begin{equation}\label{eq:reinforcementratio}
a=\frac{p(\textbf{g\textsubscript{p}})}{p(\textbf{g})} \frac{p(r|\textbf{g\textsubscript{p}},\textbf{V},\textbf{a})}{p(r|\textbf{g},\textbf{V},\textbf{a})}
{\kappa\textsuperscript{-1}}.
\end{equation}

Instead of using the ratio of likelihoods of the proposed and current models as 
%used 
in the supervised case, the ratio of episodic rewards for a model $\textbf{g}$, episodic observations $\textbf{V}$ and episodic actions $\textbf{a}$ are used: 
this is simply the scalar reward value obtained after an episode acting according to a proposed model $\textbf{g}\textsubscript{p}$ divided by the reward received when the current model $\textbf{g}$ was accepted. The prior calculation is the same as in Equation~\ref{eq:priorcalculation}. The acceptance ratio is then multiplied by a constant $\kappa\textsuperscript{-1}$ which acts a hyper-parameter that determines the extent of exploration from higher to lower reward regions on the posterior. If the acceptance ratio is less than a uniform random number generated between $0$ and $1$ then the proposed model at the first row is rejected and the row counter, $C\textsubscript{0}$, is incremented by one. A new model is then proposed at the first row. If this new model achieves a cumulative reward during the next training episode, ($p(r|\textbf{g\textsubscript{p}},\textbf{V},\textbf{a})$), such that the acceptance ratio is now greater than a uniform random number, the two models within the first rows of both arrays are accepted and then become the current models. The reward which was obtained when these current models were accepted is recorded and thereafter used as $p(r|\textbf{g},\textbf{V},\textbf{a})$ in the calculation the acceptance ratio. After training has been completed upon the algorithm reaching the final array row, actions are determined during $100$ testing episodes by calculating:
\begin{equation}\label{eq:reinforcementinference_methods}
P(a\textsubscript{new}=1|\textbf{V},\textbf{r})=\frac{1}{Tot}
\sum_{n=\beta}^{N-1} {C\textsubscript{n}\textbf{(\textbf{V}\textsubscript{new}$\cdot$\textbf{g}\textsubscript{n}})},
\end{equation}
for each array at each simulaton timestep. The summation considers only rows greater than index $\beta$, which determines the number of rows discarded to account for the burn-in period. The variable $Tot$ is the sum of all row counter values recorded after the burn-in period. The array with the largest response at each timestep determines the action taken during inference in a winner-take-all fashion. %Finally, i
It should be noted that although we have used the notation $p(r|\textbf{g},\textbf{V},\textbf{a})$, for means of consistency with the rest of the paper, this quantity is not a probability but in fact a reward.

\subsection*{Neural Network Benchmark Models}
In order to benchmark the performance of RRAM-based MCMC sampling against a state of the art machine learning approach, two neural network models were implemented using the TensorFlow python library and applied to the same tasks. Both of these models were composed of a total number of synapses equal to the number of differential conductance pairs in the memory arrays ($4,096$ in both cases).
Both models were three layer feed-forward neural networks using 32-bit floating point precision synaptic weights. The hidden layer of each model was sized such that the number of synaptic weights in the network was equal to $4,096$. The size of the first layer is consistent with the number of input features.
In the supervised neural network, the hidden and output units were logistic functions, as this is the function used in our logistic regression model. 
%The hidden layer was composed of $215$ logistic units and the output layer in the supervised learning model used two logistic units. 
The model was optimised using by minimising the categorical cross-entropy in the training data set over $100$ training epochs with the adaptive moment estimation (Adam) optimisation algorithm.
In the reinforcement learning model the hidden and output units were non-rectifying linear units since this corresponded to the row function block used in our reinforcement learning model. 
%A hidden layer of $585$ non-rectifying linear units and an output layer of two non-rectifying linear units were used. 
The parameters of the models were optimised by minimising the mean squared loss using the experience replay technique\cite{Mnih15} also using the Adam optimisation algorithm.

\subsection*{Data and Code Availability}
The Wisconsin breast cancer dataset\cite{Wolberg91} and the reinforcement learning simulation environment are publicly available. All other measured data and/or software programs used in the presentation of the paper are freely available upon request.

%\clearpage

\bibliography{bibfile}

\begin{thebibliography}{10}
\urlstyle{rm}
\expandafter\ifx\csname url\endcsname\relax
  \def\url#1{\texttt{#1}}\fi
\expandafter\ifx\csname urlprefix\endcsname\relax\def\urlprefix{URL }\fi
\expandafter\ifx\csname doiprefix\endcsname\relax\def\doiprefix{DOI: }\fi
\providecommand{\bibinfo}[2]{#2}
\providecommand{\eprint}[2][]{\url{#2}}

\bibitem{Shi16}
\bibinfo{author}{{Shi}, W.}, \bibinfo{author}{{Cao}, J.},
  \bibinfo{author}{{Zhang}, Q.}, \bibinfo{author}{{Li}, Y.} \&
  \bibinfo{author}{{Xu}, L.}
\newblock \bibinfo{journal}{\bibinfo{title}{Edge computing: Vision and
  challenges}}.
\newblock {\emph{\JournalTitle{IEEE Internet of Things Journal}}}
  \textbf{\bibinfo{volume}{3}}, \bibinfo{pages}{637--646}
  (\bibinfo{year}{2016}).

\bibitem{vonNeumann93}
\bibinfo{author}{{von Neumann}, J.}
\newblock \bibinfo{journal}{\bibinfo{title}{First draft of a report on the
  edvac}}.
\newblock {\emph{\JournalTitle{IEEE Annals of the History of Computing}}}
  \textbf{\bibinfo{volume}{15}}, \bibinfo{pages}{27--75}
  (\bibinfo{year}{1993}).

\bibitem{LeCun15}
\bibinfo{author}{LeCun, Y.}, \bibinfo{author}{Bengio, Y.} \&
  \bibinfo{author}{Hinton, G.}
\newblock \bibinfo{journal}{\bibinfo{title}{Deep learning}}.
\newblock {\emph{\JournalTitle{Nature}}} \textbf{\bibinfo{volume}{521}},
  \bibinfo{pages}{436--44} (\bibinfo{year}{2015}).

\bibitem{Strubell19}
\bibinfo{author}{Strubell, E.}, \bibinfo{author}{Ganesh, A.} \&
  \bibinfo{author}{Mccallum, A.}
\newblock \bibinfo{title}{Energy and policy considerations for deep learning in
  nlp}.
\newblock In \emph{\bibinfo{booktitle}{In the 57th Annual Meeting of the
  Association for Computational Linguistics (ACL). Florence, Italy. July 2019}}
  (\bibinfo{year}{2019}).

\bibitem{Li16}
\bibinfo{author}{{Li}, D.}, \bibinfo{author}{{Chen}, X.},
  \bibinfo{author}{{Becchi}, M.} \& \bibinfo{author}{{Zong}, Z.}
\newblock \bibinfo{title}{Evaluating the energy efficiency of deep
  convolutional neural networks on cpus and gpus}.
\newblock In \emph{\bibinfo{booktitle}{2016 IEEE International Conferences on
  Big Data and Cloud Computing (BDCloud), Social Computing and Networking
  (SocialCom), Sustainable Computing and Communications (SustainCom)
  (BDCloud-SocialCom-SustainCom)}}, \bibinfo{pages}{477--484}
  (\bibinfo{year}{2016}).

\bibitem{Chua71}
\bibinfo{author}{{Chua}, L.}
\newblock \bibinfo{journal}{\bibinfo{title}{Memristor-the missing circuit
  element}}.
\newblock {\emph{\JournalTitle{IEEE Transactions on Circuit Theory}}}
  \textbf{\bibinfo{volume}{18}}, \bibinfo{pages}{507--519}
  (\bibinfo{year}{1971}).

\bibitem{Strukov08}
\bibinfo{author}{Strukov, D.}, \bibinfo{author}{Snider, G.},
  \bibinfo{author}{Stewart, D.} \& \bibinfo{author}{Williams, S.}
\newblock \bibinfo{journal}{\bibinfo{title}{The missing memristor found}}.
\newblock {\emph{\JournalTitle{Nature}}} \textbf{\bibinfo{volume}{453}},
  \bibinfo{pages}{80--3} (\bibinfo{year}{2008}).

\bibitem{Xia19}
\bibinfo{author}{Xia, Q.} \& \bibinfo{author}{Yang, J.~J.}
\newblock \bibinfo{journal}{\bibinfo{title}{Memristive crossbar arrays for
  brain-inspired computing}}.
\newblock {\emph{\JournalTitle{Nature Materials}}}
  \textbf{\bibinfo{volume}{18}}, \bibinfo{pages}{309--323}
  (\bibinfo{year}{2019}).

\bibitem{Ambrogio18}
\bibinfo{author}{Ambrogio, S.} \emph{et~al.}
\newblock \bibinfo{journal}{\bibinfo{title}{Equivalent-accuracy accelerated
  neural-network training using analogue memory}}.
\newblock {\emph{\JournalTitle{Nature}}} \textbf{\bibinfo{volume}{558}}
  (\bibinfo{year}{2018}).

\bibitem{Prezioso15}
\bibinfo{author}{Prezioso, M.} \emph{et~al.}
\newblock \bibinfo{journal}{\bibinfo{title}{Training and operation of an
  integrated neuromorphic network based on metal-oxide memristors}}.
\newblock {\emph{\JournalTitle{Nature}}} \textbf{\bibinfo{volume}{521}}
  (\bibinfo{year}{2014}).

\bibitem{Wang19}
\bibinfo{author}{Wang, Z.} \emph{et~al.}
\newblock \bibinfo{journal}{\bibinfo{title}{Reinforcement learning with
  analogue memristor arrays}}.
\newblock {\emph{\JournalTitle{Nature Electronics}}}
  \textbf{\bibinfo{volume}{2}} (\bibinfo{year}{2019}).

\bibitem{Miao18}
\bibinfo{author}{Hu, M.} \emph{et~al.}
\newblock \bibinfo{journal}{\bibinfo{title}{Memristor-based analog computation
  and neural network classification with a dot product engine}}.
\newblock {\emph{\JournalTitle{Advanced Materials}}}
  \textbf{\bibinfo{volume}{30}}, \bibinfo{pages}{1705914}
  (\bibinfo{year}{2018}).

\bibitem{Li18}
\bibinfo{author}{Li, C.} \emph{et~al.}
\newblock \bibinfo{journal}{\bibinfo{title}{Efficient and self-adaptive in-situ
  learning in multilayer memristor neural networks}}.
\newblock {\emph{\JournalTitle{Nature Communications}}}
  \textbf{\bibinfo{volume}{9}} (\bibinfo{year}{2018}).

\bibitem{Wong10}
\bibinfo{author}{{Wong}, H. .~P.} \emph{et~al.}
\newblock \bibinfo{journal}{\bibinfo{title}{Phase change memory}}.
\newblock {\emph{\JournalTitle{Proceedings of the IEEE}}}
  \textbf{\bibinfo{volume}{98}}, \bibinfo{pages}{2201--2227}
  (\bibinfo{year}{2010}).

\bibitem{Chappert07}
\bibinfo{author}{Chappert, C.}, \bibinfo{author}{Fert, A.} \&
  \bibinfo{author}{Dau, F.}
\newblock \bibinfo{journal}{\bibinfo{title}{The emergence of spin electronics
  in data storage}}.
\newblock {\emph{\JournalTitle{Nature materials}}}
  \textbf{\bibinfo{volume}{6}}, \bibinfo{pages}{813--23}
  (\bibinfo{year}{2007}).

\bibitem{Liu12}
\bibinfo{author}{Liu, Q.} \emph{et~al.}
\newblock \bibinfo{journal}{\bibinfo{title}{Resistive switching: Real-time
  observation on dynamic growth/dissolution of conductive filaments in
  oxide-electrolyte-based reram (adv. mater. 14/2012)}}.
\newblock {\emph{\JournalTitle{Advanced Materials}}}
  \textbf{\bibinfo{volume}{24}}, \bibinfo{pages}{1774--1774}
  (\bibinfo{year}{2012}).

\bibitem{Beck00}
\bibinfo{author}{Beck, A.}, \bibinfo{author}{Bednorz, J.~G.},
  \bibinfo{author}{Gerber, C.}, \bibinfo{author}{Rossel, C.} \&
  \bibinfo{author}{Widmer, D.}
\newblock \bibinfo{journal}{\bibinfo{title}{Reproducible switching effect in
  thin oxide films for memory applications}}.
\newblock {\emph{\JournalTitle{Applied Physics Letters}}}
  \textbf{\bibinfo{volume}{77}}, \bibinfo{pages}{139--141}
  (\bibinfo{year}{2000}).

\bibitem{Gokmen17}
\bibinfo{author}{Gokmen, T.}, \bibinfo{author}{Onen, M.} \&
  \bibinfo{author}{Haensch, W.}
\newblock \bibinfo{journal}{\bibinfo{title}{Training deep convolutional neural
  networks with resistive cross-point devices}}.
\newblock {\emph{\JournalTitle{Frontiers in Neuroscience}}}
  \textbf{\bibinfo{volume}{11}}, \bibinfo{pages}{538} (\bibinfo{year}{2017}).

\bibitem{Burr15}
\bibinfo{author}{{Burr}, G.~W.} \emph{et~al.}
\newblock \bibinfo{journal}{\bibinfo{title}{Experimental demonstration and
  tolerancing of a large-scale neural network (165 000 synapses) using
  phase-change memory as the synaptic weight element}}.
\newblock {\emph{\JournalTitle{IEEE Transactions on Electron Devices}}}
  \textbf{\bibinfo{volume}{62}}, \bibinfo{pages}{3498--3507}
  (\bibinfo{year}{2015}).

\bibitem{Garbin15}
\bibinfo{author}{{Garbin}, D.} \emph{et~al.}
\newblock \bibinfo{journal}{\bibinfo{title}{Hfo2-based oxram devices as
  synapses for convolutional neural networks}}.
\newblock {\emph{\JournalTitle{IEEE Transactions on Electron Devices}}}
  \textbf{\bibinfo{volume}{62}}, \bibinfo{pages}{2494--2501}
  (\bibinfo{year}{2015}).

\bibitem{Sebastian15}
\bibinfo{author}{Sebastian, A.}, \bibinfo{author}{Krebs, D.},
  \bibinfo{author}{Gallo, M.~L.}, \bibinfo{author}{Pozidis, H.} \&
  \bibinfo{author}{Eleftheriou, E.}
\newblock \bibinfo{journal}{\bibinfo{title}{A collective relaxation model for
  resistance drift in phase change memory cells}}.
\newblock {\emph{\JournalTitle{2015 IEEE International Reliability Physics
  Symposium}}} \bibinfo{pages}{MY.5.1--MY.5.6} (\bibinfo{year}{2015}).

\bibitem{Guan12}
\bibinfo{author}{{Guan}, X.}, \bibinfo{author}{{Yu}, S.} \&
  \bibinfo{author}{{Wong}, H. .~P.}
\newblock \bibinfo{journal}{\bibinfo{title}{On the switching parameter
  variation of metal-oxide rram—part i: Physical modeling and simulation
  methodology}}.
\newblock {\emph{\JournalTitle{IEEE Transactions on Electron Devices}}}
  \textbf{\bibinfo{volume}{59}}, \bibinfo{pages}{1172--1182}
  (\bibinfo{year}{2012}).

\bibitem{Yu12pt2}
\bibinfo{author}{{Yu}, S.}, \bibinfo{author}{{Guan}, X.} \&
  \bibinfo{author}{{Wong}, H. .~P.}
\newblock \bibinfo{journal}{\bibinfo{title}{On the switching parameter
  variation of metal oxide rram—part ii: Model corroboration and device
  design strategy}}.
\newblock {\emph{\JournalTitle{IEEE Transactions on Electron Devices}}}
  \textbf{\bibinfo{volume}{59}}, \bibinfo{pages}{1183--1188}
  (\bibinfo{year}{2012}).

\bibitem{Sidler16}
\bibinfo{author}{{Sidler}, S.} \emph{et~al.}
\newblock \bibinfo{title}{Large-scale neural networks implemented with
  non-volatile memory as the synaptic weight element: Impact of conductance
  response}.
\newblock In \emph{\bibinfo{booktitle}{2016 46th European Solid-State Device
  Research Conference (ESSDERC)}}, \bibinfo{pages}{440--443}
  (\bibinfo{year}{2016}).

\bibitem{Bennett19}
\bibinfo{author}{{Bennett}, C.~H.}, \bibinfo{author}{{Garland}, D.},
  \bibinfo{author}{{Jacobs-Gedrim}, R.~B.}, \bibinfo{author}{{Agarwal}, S.} \&
  \bibinfo{author}{{Marinella}, M.~J.}
\newblock \bibinfo{title}{Wafer-scale taox device variability and implications
  for neuromorphic computing applications}.
\newblock In \emph{\bibinfo{booktitle}{2019 IEEE International Reliability
  Physics Symposium (IRPS)}}, \bibinfo{pages}{1--4} (\bibinfo{year}{2019}).

\bibitem{Agarwal16}
\bibinfo{author}{{Agarwal}, S.} \emph{et~al.}
\newblock \bibinfo{title}{Resistive memory device requirements for a neural
  algorithm accelerator}.
\newblock In \emph{\bibinfo{booktitle}{2016 International Joint Conference on
  Neural Networks (IJCNN)}}, \bibinfo{pages}{929--938} (\bibinfo{year}{2016}).

\bibitem{Nandakumar18}
\bibinfo{author}{{Nandakumar}, S.~R.} \emph{et~al.}
\newblock \bibinfo{title}{Mixed-precision architecture based on computational
  memory for training deep neural networks}.
\newblock In \emph{\bibinfo{booktitle}{2018 IEEE International Symposium on
  Circuits and Systems (ISCAS)}}, \bibinfo{pages}{1--5} (\bibinfo{year}{2018}).

\bibitem{Boybat17}
\bibinfo{author}{Boybat, I.} \emph{et~al.}
\newblock \bibinfo{journal}{\bibinfo{title}{Neuromorphic computing with
  multi-memristive synapses}}.
\newblock {\emph{\JournalTitle{Nature Communications}}}
  \textbf{\bibinfo{volume}{9}} (\bibinfo{year}{2017}).

\bibitem{Serb16}
\bibinfo{author}{Serb, A.} \emph{et~al.}
\newblock \bibinfo{journal}{\bibinfo{title}{Unsupervised learning in
  probabilistic neural networks with multi-state metal-oxide memristive
  synapses}}.
\newblock {\emph{\JournalTitle{Nature Communications}}}
  \textbf{\bibinfo{volume}{7}}, \bibinfo{pages}{12611} (\bibinfo{year}{2016}).

\bibitem{Querlioz13}
\bibinfo{author}{{Querlioz}, D.}, \bibinfo{author}{{Bichler}, O.},
  \bibinfo{author}{{Dollfus}, P.} \& \bibinfo{author}{{Gamrat}, C.}
\newblock \bibinfo{journal}{\bibinfo{title}{Immunity to device variations in a
  spiking neural network with memristive nanodevices}}.
\newblock {\emph{\JournalTitle{IEEE Transactions on Nanotechnology}}}
  \textbf{\bibinfo{volume}{12}}, \bibinfo{pages}{288--295}
  (\bibinfo{year}{2013}).

\bibitem{Dalgaty19}
\bibinfo{author}{Dalgaty, T.} \emph{et~al.}
\newblock \bibinfo{journal}{\bibinfo{title}{Hybrid neuromorphic circuits
  exploiting non-conventional properties of rram for massively parallel local
  plasticity mechanisms}}.
\newblock {\emph{\JournalTitle{APL Materials}}} \textbf{\bibinfo{volume}{7}},
  \bibinfo{pages}{081125} (\bibinfo{year}{2019}).

\bibitem{Chen15}
\bibinfo{author}{{Chen}, A.}
\newblock \bibinfo{journal}{\bibinfo{title}{Utilizing the variability of
  resistive random access memory to implement reconfigurable physical
  unclonable functions}}.
\newblock {\emph{\JournalTitle{IEEE Electron Device Letters}}}
  \textbf{\bibinfo{volume}{36}}, \bibinfo{pages}{138--140}
  (\bibinfo{year}{2015}).

\bibitem{Balatti15}
\bibinfo{author}{{Balatti}, S.}, \bibinfo{author}{{Ambrogio}, S.},
  \bibinfo{author}{{Wang}, Z.} \& \bibinfo{author}{{Ielmini}, D.}
\newblock \bibinfo{journal}{\bibinfo{title}{True random number generation by
  variability of resistive switching in oxide-based devices}}.
\newblock {\emph{\JournalTitle{IEEE Journal on Emerging and Selected Topics in
  Circuits and Systems}}} \textbf{\bibinfo{volume}{5}},
  \bibinfo{pages}{214--221} (\bibinfo{year}{2015}).

\bibitem{vodenicarevic2017low}
\bibinfo{author}{Vodenicarevic, D.} \emph{et~al.}
\newblock \bibinfo{journal}{\bibinfo{title}{Low-energy truly random number
  generation with superparamagnetic tunnel junctions for unconventional
  computing}}.
\newblock {\emph{\JournalTitle{Physical Review Applied}}}
  \textbf{\bibinfo{volume}{8}}, \bibinfo{pages}{054045} (\bibinfo{year}{2017}).

\bibitem{shim2017stochastic}
\bibinfo{author}{Shim, Y.}, \bibinfo{author}{Chen, S.},
  \bibinfo{author}{Sengupta, A.} \& \bibinfo{author}{Roy, K.}
\newblock \bibinfo{journal}{\bibinfo{title}{Stochastic spin-orbit torque
  devices as elements for bayesian inference}}.
\newblock {\emph{\JournalTitle{Scientific reports}}}
  \textbf{\bibinfo{volume}{7}}, \bibinfo{pages}{14101} (\bibinfo{year}{2017}).

\bibitem{faria2018implementing}
\bibinfo{author}{Faria, R.}, \bibinfo{author}{Camsari, K.~Y.} \&
  \bibinfo{author}{Datta, S.}
\newblock \bibinfo{journal}{\bibinfo{title}{Implementing bayesian networks with
  embedded stochastic mram}}.
\newblock {\emph{\JournalTitle{AIP Advances}}} \textbf{\bibinfo{volume}{8}},
  \bibinfo{pages}{045101} (\bibinfo{year}{2018}).

\bibitem{mizrahi2018neural}
\bibinfo{author}{Mizrahi, A.} \emph{et~al.}
\newblock \bibinfo{journal}{\bibinfo{title}{Neural-like computing with
  populations of superparamagnetic basis functions}}.
\newblock {\emph{\JournalTitle{Nature communications}}}
  \textbf{\bibinfo{volume}{9}}, \bibinfo{pages}{1533} (\bibinfo{year}{2018}).

\bibitem{camsari2017stochastic}
\bibinfo{author}{Camsari, K.~Y.}, \bibinfo{author}{Faria, R.},
  \bibinfo{author}{Sutton, B.~M.} \& \bibinfo{author}{Datta, S.}
\newblock \bibinfo{journal}{\bibinfo{title}{Stochastic p-bits for invertible
  logic}}.
\newblock {\emph{\JournalTitle{Physical Review X}}}
  \textbf{\bibinfo{volume}{7}}, \bibinfo{pages}{031014} (\bibinfo{year}{2017}).

\bibitem{borders2019integer}
\bibinfo{author}{Borders, W.~A.} \emph{et~al.}
\newblock \bibinfo{journal}{\bibinfo{title}{Integer factorization using
  stochastic magnetic tunnel junctions}}.
\newblock {\emph{\JournalTitle{Nature}}} \textbf{\bibinfo{volume}{573}},
  \bibinfo{pages}{390--393} (\bibinfo{year}{2019}).

\bibitem{Hastings70}
\bibinfo{author}{Hastings, W.~K.}
\newblock \bibinfo{journal}{\bibinfo{title}{Monte carlo sampling methods using
  markov chains and their applications}}.
\newblock {\emph{\JournalTitle{Biometrika}}} \textbf{\bibinfo{volume}{57}},
  \bibinfo{pages}{97--109} (\bibinfo{year}{1970}).

\bibitem{Ghahramani15}
\bibinfo{author}{Ghahramani, Z.}
\newblock \bibinfo{journal}{\bibinfo{title}{Probabilistic machine learning and
  artificial intelligence}}.
\newblock {\emph{\JournalTitle{Nature}}} \textbf{\bibinfo{volume}{521}},
  \bibinfo{pages}{452--9} (\bibinfo{year}{2015}).

\bibitem{Grossi16}
\bibinfo{author}{{Grossi}, A.} \emph{et~al.}
\newblock \bibinfo{title}{Fundamental variability limits of filament-based
  rram}.
\newblock In \emph{\bibinfo{booktitle}{2016 IEEE International Electron Devices
  Meeting (IEDM)}}, \bibinfo{pages}{4.7.1--4.7.4} (\bibinfo{year}{2016}).

\bibitem{Ielmini11}
\bibinfo{author}{{Ielmini}, D.}
\newblock \bibinfo{journal}{\bibinfo{title}{Modeling the universal set/reset
  characteristics of bipolar rram by field- and temperature-driven filament
  growth}}.
\newblock {\emph{\JournalTitle{IEEE Transactions on Electron Devices}}}
  \textbf{\bibinfo{volume}{58}}, \bibinfo{pages}{4309--4317}
  (\bibinfo{year}{2011}).

\bibitem{Wolberg91}
\bibinfo{author}{Wolberg, W.~H.} \& \bibinfo{author}{Mangasarian, O.~L.}
\newblock \bibinfo{journal}{\bibinfo{title}{Multisurface method of pattern
  separation for medical diagnosis applied to breast cytology.}}
\newblock {\emph{\JournalTitle{Proceedings of the National Academy of
  Sciences}}} \textbf{\bibinfo{volume}{87}}, \bibinfo{pages}{9193--9196}
  (\bibinfo{year}{1990}).

\bibitem{Sutton98}
\bibinfo{author}{Sutton, R.~S.} \& \bibinfo{author}{Barto, A.~G.}
\newblock \emph{\bibinfo{title}{Introduction to Reinforcement Learning}}
  (\bibinfo{publisher}{MIT Press}, \bibinfo{year}{1998}).

\bibitem{Hoffman07}
\bibinfo{author}{Hoffman, M.}, \bibinfo{author}{Doucet, A.},
  \bibinfo{author}{Freitas, N.~d.} \& \bibinfo{author}{Jasra, A.}
\newblock \bibinfo{title}{Trans-dimensional mcmc for bayesian policy learning}.
\newblock In \emph{\bibinfo{booktitle}{Proceedings of the 20th International
  Conference on Neural Information Processing Systems}}, NIPS’07,
  \bibinfo{pages}{665–672} (\bibinfo{publisher}{Curran Associates Inc.},
  \bibinfo{address}{Red Hook, NY, USA}, \bibinfo{year}{2007}).

\bibitem{Barto83}
\bibinfo{author}{{Barto}, A.~G.}, \bibinfo{author}{{Sutton}, R.~S.} \&
  \bibinfo{author}{{Anderson}, C.~W.}
\newblock \bibinfo{journal}{\bibinfo{title}{Neuronlike adaptive elements that
  can solve difficult learning control problems}}.
\newblock {\emph{\JournalTitle{IEEE Transactions on Systems, Man, and
  Cybernetics}}} \textbf{\bibinfo{volume}{SMC-13}}, \bibinfo{pages}{834--846}
  (\bibinfo{year}{1983}).

\bibitem{Mnih15}
\bibinfo{author}{Mnih, V.} \emph{et~al.}
\newblock \bibinfo{journal}{\bibinfo{title}{Human-level control through deep
  reinforcement learning}}.
\newblock {\emph{\JournalTitle{Nature}}} \textbf{\bibinfo{volume}{518}},
  \bibinfo{pages}{529--33} (\bibinfo{year}{2015}).

\bibitem{Liu95}
\bibinfo{author}{Liu, H.} \& \bibinfo{author}{Setiono, R.}
\newblock \bibinfo{journal}{\bibinfo{title}{Chi2: feature selection and
  discretization of numeric attributes}}.
\newblock {\emph{\JournalTitle{Proceedings of 7th IEEE International Conference
  on Tools with Artificial Intelligence}}} \bibinfo{pages}{388--391}
  (\bibinfo{year}{1995}).

\end{thebibliography}
%TC:endignore
\end{document}